\documentclass[useAMS,usenatbib]{mn2e}
\usepackage{graphicx}
\usepackage{color}

\title[Trojan satellites as a plasma source]
{On the ultraviolet anomalies of the WASP-12 and HD~189733 systems: Trojan satellites as a plasma source}
\author[K.~G.~Kislyakova et al.]
  {K.~G.~Kislyakova$^{1}$, E.~Pilat-Lohinger$^{2}$, B.~Funk$^{2}$, H.~Lammer$^{1}$, L.~Fossati$^{1}$, 
  \newauthor
  S.~Eggl$^{3}$, R.~Schwarz$^{2}$, M.~Y.~Boudjada$^{1}$, and N.~V.~Erkaev$^{4,5}$ \\
$^{1}$Space Research Institute, Austrian Academy of Sciences, Graz, Austria; kristina.kislyakova@oeaw.ac.at\\
$^{2}$University of Vienna, Department of Astrophysics, Vienna, Austria \\
$^{3}$IMCCE Observatoire de Paris, Univ. Lille 1, UPMC, 77 Avenue Denfert-Rochereau, 75014 Paris, France  \\
$^{4}$Institute of Computational Modelling, Siberian Division of Russian Academy of Sciences, Krasnoyarsk, Russian Federation \\
$^{5}$Siberian Federal University, Krasnoyarsk, Russian Federation }
\date{Released 2015 Xxxxx XX}

\pagerange{\pageref{firstpage}--\pageref{lastpage}} \pubyear{2015}

\def\LaTeX{L\kern-.36em\raise.3ex\hbox{a}\kern-.15em
    T\kern-.1667em\lower.7ex\hbox{E}\kern-.125emX}

\begin{document}

\label{firstpage}

\maketitle

\begin{abstract}
We suggest an additional possible plasma source to explain part of the phenomena observed for the transiting hot Jupiters WASP-12b and HD 189733b in their ultraviolet (UV) light curves. In the proposed scenario, material outgasses from the molten surface of Trojan satellites on tadpole orbits near the Lagrange points L$_4$ and L$_5$. We show that the temperature at the orbital location of WASP-12b is high enough to melt the surface of rocky bodies and to form shallow lava oceans on them. In case of WASP-12b, this leads to the release of elements such as Mg and Ca, which are expected to surround the system. The predicted Mg and Ca outgassing rates from two Io-sized WASP-12b Trojans are $\approx 2.2 \times 10^{27}$ s$^{-1}$ and $\approx 2.2 \times 10^{26}$ s$^{-1}$, respectively. Trojan outgassing can lead to the apparent lack of emission in Mg{\sc ii}\,h\&k and Ca{\sc ii}\,H\&K line cores of WASP-12. For HD 189733b, the mechanism is only marginally possible due to the lower temperature. This may be one of the reasons that couldn't explain the early ingress of HD 189733b observed in the far-UV (FUV) C{\sc ii} doublet due to absence of carbon within elements outgassed by molten lava. We investigate the long-term stability region of WASP-12b and HD 189733b in case of planar and inclined motion of these satellites and show that unlike the classical exomoons orbiting the planet, Io-sized Trojans can be stable for the whole systems life time. 
\end{abstract}


\section{Introduction}
\label{sec_intro}

\begin{figure}
  \begin{center}
		\includegraphics[width=1.0\columnwidth]{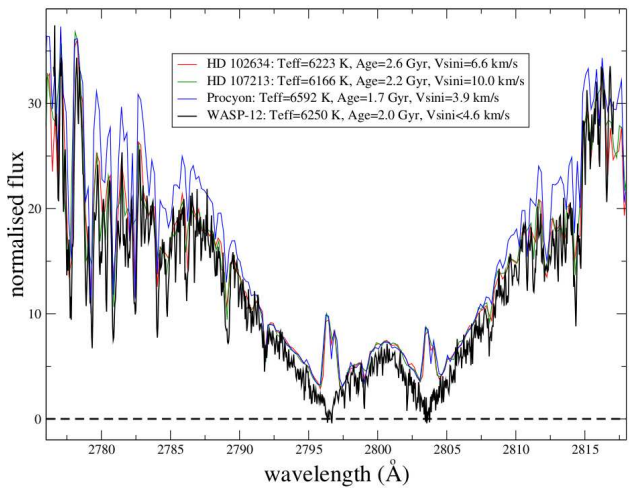}
        \caption{NUV coverage of the cores of the Mg{\sc ii} resonance lines compared with those of three other stars \citep{Haswell12}. The cores of the WASP-12 lines show no emission in comparison to other stars on the same activity level. The remarkable appearance of Mg{\sc ii} resonance line cores of WASP-12b is believed to be caused by absorbing material located inside the WASP-12 system.}
     \label{f_haswell}
  \end{center}
\end{figure}

Hot Jupiters, Jupiter-mass exoplanets orbiting closer than 0.1~AU from their host stars, are a natural laboratory to study the effects and interaction processes, which are much weaker in the Solar system. Due to high stellar irradiance and gravity at close orbital distance, the star-planet interaction is very intense in such systems. WASP-12b is an extreme hot Jupiter with an equilibrium temperature exceeding 2500~K orbiting its host star only two stellar radii from its surface. WASP-12b is a transiting planet which has been observed at different wavelengths including ultraviolet. These observations revealed a number of interesting features, namely, early ingress observed in near ultra-violet (NUV) in comparison to the optical transit and complete suppression of NUV Mg{\sc ii}\,h\&k and optical Ca{\sc ii}\,H\&K emission line cores (\citealp{Fossati10,Haswell12}; see Fig.~\ref{f_haswell}). Several explanations have been proposed for these observations. For early ingress, the proposed hypothesis included absorption in the bow shock ahead of the planet with the stand-off distance determined by the planetary magnetic moment \citep{Vidotto10}, and evaporation of atmospheric material from the planet \citep{Fossati10,Haswell12,Bisikalo13,Bisikalo15}. The apparent lack of emission in Mg{\sc ii}\,h\&k and Ca{\sc ii}\,H\&K line cores of WASP-12 was originally explained by the presence of absorbing material evaporated from the planet \citep{Haswell12,Fossati13}.

WASP-12b is not the only hot Jupiter which revealed the early ingress observed at UV wavelengths. Another example is the hot Jupiter HD~189733b, for which an early ingress in FUV C{\sc ii} doublet has been detected \citep{Ben-Jaffel14}. Note that \citet{Haswell12} looked for an early ingress for HD~189733b using exactly the same data later analysed by \citet{Ben-Jaffel14} without finding any.

 Recently, \citet{Ben-Jaffel14} proposed a new hypothesis that the early ingress observed for the transits of the hot Jupiters WASP-12b and HD~189733b is caused by transit of exomoon plasma tori, similar to the one produced by the Jovian satellite Io. The satellites of WASP-12b and HD~189733b were assumed to orbit at approximately 6 and 16 planetary radii from the planets' centers, respectively. This hypothesis offers new insights into these exoplanetary systems and is widely cited in the literature {(e.g., \citealp{Heller14b,Fossati14,Ballester15,Simon15}). However, it did not account for some special conditions in hot Jupiter systems.

 Although there are no reasons to question the existence of exomoons in general, the moons of close-in exoplanets must have tight orbits. Moreover, one has to place strict limits to moons' size and mass. A lower limit is necessary due to the strong stellar radiation in this area and an upper limit is required for the dynamical stability \citep{Weidner10}. In hot Jupiter systems, the satellites are endangered by several mechanisms: {\it i}) they can be destroyed by tidal forces if they move too close to the exoplanet (i.e. inside the Roche limit), {\it ii}) they become gravitationally unbound from the planet if they are outside its Hill radius, and {\it iii}) the orbits of too massive satellites are disturbed by tidal forces, which move them towards the planet and eventually destroy them when the satellite reaches the Roche limit \citep{Barnes02,Weidner10,Domingos06}. Therefore, even for favorable orbits, the sizes and lifetimes of the exomoons are highly restricted and incomparable to those of the Solar system moons.

 However, one can not exclude the existence of co-orbital satellites in the stable $L_4$ and $L_5$ Lagrangian points of the exoplanet-star system orbiting in 1:1 resonance with the giant planet. In the Solar system, numerous Trojan satellites have been discovered at the Lagrangian equilibrium points L$_4$ and L$_5$ of different planets (Earth, Mars, Jupiter, Uranus and Neptune, \citealp{Jewitt00,Jewitt07,Schwarz12,Galiazzo14}). 
 
Detecting moons and Trojans requires extremely high precision and stability photometry. At present neither moons nor Trojans have been reliably detected in exoplanetary systems. \citet{Placek15} recently announced the detection of a Trojan object in the Kepler 91 system, where a Jupiter-mass planet orbits a sun-like star in 0.072~AU. The authors note that the discovery might be a false positive, however, if confirmed, it would be the first evidence of a Trojan object of a close-in exoplanet proving that hot Jupiters can host Trojans like the giant planets in our Solar system.

 In this study, we propose Trojan satellites as an outgassing source in the WASP-12 and HD~189733 systems, discuss the formation of lava oceans on their surfaces and estimate possible mineral outgassing rates. We consider two single Trojan satellites of the size of Io in $L_4$ and $L_5$ and show that they are stable for system's life time. Additionally, we study also two Trojan swarms consisting of 100 objects each with a radius of 100~km. But we have to point out that two single Io-sized bodies are probably more stable on longer time scales. Figure~\ref{f_sketch} summarizes the considered orbital configurations. 
 
We also re-examine Io-type volcanism as an outgassing source for the observed materials. \citet{Ben-Jaffel14} proposed Io-type volcanic outgassing of such elements as Mg and C at a rate of $\approx 10^{28}$ and $\approx 10^{29}$ atoms per second, respectively. In the Jovian system, the extreme volcanism and plasma production of Io is explained by Io's eccentricity, which increases the tidal heating. The high eccentricity is supported by two other Galilean moons, Europa and Ganymede \citep{Peale79}. In systems hosting hot Jupiters, this mechanism of interior heating and plasma production can not operate due to the absence of a family of big classical moons. Also, Mg and C are not present in the materials outgassed by Io's volcanoes. We suppose that in hot Jupiter systems, plasma can be produced by other means, namely, it can originate from the sparse gaseous atmosphere surrounding Trojan satellites. Plasma can be outgassed from the molten rocky surfaces of Trojans, similarly to the formation of silicate atmospheres of hot close-in rocky planets \citep{Schaefer09,Mura11,Miguel11,Ito15}. This mechanism of atmosphere formation and escape was studied in detail for CoRoT-7b by \citet{Mura11}. Here we apply it to possible Trojan satellites in WASP-12 and HD~189733 systems and estimate the outgassing rates. 

\begin{figure}
\includegraphics[width=1.0\columnwidth]{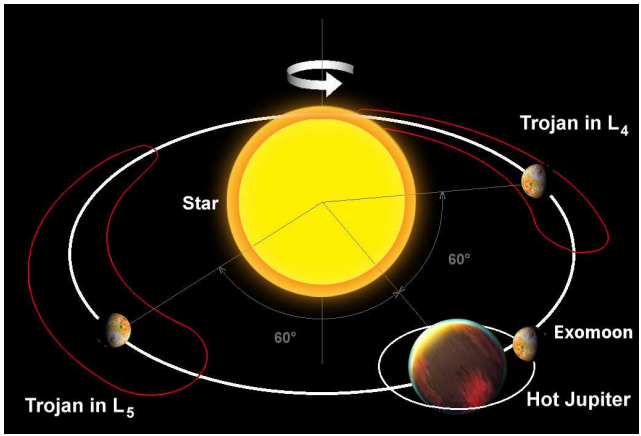}
 \vspace{3.5cm}
 \caption{A sketch illustrating orbital configurations considered in the present study. \textit{i}) A classical exomoon (a satellite orbiting a planet), which we show to be unstable in most of the cases; \textit{ii}) Trojan satellites orbiting on tadpole orbits in the stability region around the L$_4$ and L$_5$ Lagrange points (orbiting in 1:1 resonance with the planet with stable orbits possible everywhere within the area marked by the red line; also inclined orbits are possible).}
\label{f_sketch}
\end{figure}

This paper is organized as follows. In Section~\ref{sec_stability}, we calculate the possible stable orbits of Trojan satellites in the WASP-12 and HD~189733 systems, the size of the stability regions and consider the photogravitational stability. In Section~\ref{sec_outgassing}, we estimate the outgassing rates from molten lava oceans on the surfaces of Trojan satellites. In Section~\ref{sec_torus}, we discuss the influence that outgassed and ionized material could have on the observed stellar activity of WASP-12 and compare it to the transit observations of WASP-12b. In Section~\ref{sec_Hill}, we apply the general stability criteria of classical (exo)moons to systems hosting close-in giant planets to check if and where classical exomoons can be present in the WASP-12 and HD~189733 systems. Finally, Section~\ref{sec_conclusions} summarizes our conclusions.

\section{Orbital stability}
\label{sec_stability}

\subsection{Possible origin of Trojan satellites}

The process of formation and migration of hot Jupiters (and, therefore, their Trojans) has been a matter of debate since the discovery of 51-Peg b, and it is still not firmly established. Several mechanisms have been proposed: planet-planet scattering \citep{Rasio96}, Kozai-Likov cycles \citep{Kozai62,Lidov62}, or high eccentricity migration on the disk \citep{Lin96}. All of them imply different fates for possible additional companions in the system before the Jupiter-sized object moved to its current orbit. In this work, we want to check whether evaporating Trojans can lead to the peculiar absorption signals detected at ultraviolet wavelengths and we do not aim at the study of the origin, capture, or migration of Trojans. However, for completeness, we summarize here below the most recent and relevant studies on this topic.

Theoretical studies predict that Trojans are a by-product of planet formation and evolution. There are two ways to form Trojan objects (as well as exomoons): \textit{in situ} formation (followed by possible migration to hot Jupiter's final position) or capture. 
 \citet{Beauge07} showed that a big Trojan satellite, with a mass never exceeding $0.15 - 0.3 M_{\rm Earth}$, can be formed \textit{in situ} from the initial Trojan population containing 300--1000 bodies. This makes the existence of one or two (in $L_4$ and $L_5$) Io-mass satellites possible. \citet{Beauge07} also studied the type II migration of a giant planet assuming that its semimajor axis decayed on a time scale of $10^5-10^7$ years. Their results showed that the Trojan satellite migrated together with the planet, and that the tadpole configuration close to L$_4$ was maintained. \citet{Moldovan10} have investigated the fate of a possible Trojan asteroid swarm composed of many objects following the predicted migration path of the hot Jupiter HD209458b. According to their results, some objects have survived migration, although the swarm was depleted.

Numerical simulations of chaotic capture of Trojans in the outer Solar system \citep{Morbidelli05,Nesvorny09} showed that capture happens with a probability of about 10$^{-5}$ for Jupiter and 10$^{-4}$ for Neptune. \citet{Schwarz12} investigated the capture mechanism in the inner solar system for Near Earth Asteroids, where the capture probability is between 10$^{-2}$ and 10$^{-3}$ for Venus, Earth and Mars. These results were confirmed for mainbelt asteroids by \citet{Galiazzo14}. The estimates for Solar system giants were based on the Nice model of late stage planet formation. In exoplanetary systems, the capture probabilities should depend on the architecture of a particular system. This was shown by \citet{Podlewska-Gaca14} who studied the possibility of a migration-induced resonance locking in systems containing three planets, namely an Earth-like planet, a super-Earth and a gas giant with one Jupiter mass and have found that that the super-Earth planet temporarily captures the Earth-like planet into a co-orbital motion.

It has been proven by high spatial resolution images that WASP-12b orbits the primary star of a hierarchical triple star system \citep{Bechter14}. Two distant M3V companions co-orbit the hot Jupiter planet host and are separated from each other by approximately 21~AU. The angular separation between WASP-12~A and WASP-12~BC is about 1'', which corresponds to a distance of $\le 300$~AU \citep{Crossfield12}. This distance is much larger than the semi-major axis of WASP-12b, which is only 0.023~AU. The K0V star HD~189733 is a binary system with a secondary star being a mid-M dwarf with a projected separation of about 216~AU from the primary \citep{Bakos06}.

In both systems, the perturbing star (or star system) is at a distance of some hundreds AU. Even if we consider an extremely high eccentricity of the binary of 0.9 the stability region around both host-stars would reach up to a distance of more than 50~AU. This illustrates that stable motion is possible up to Pluto's distance in the solar system. Moreover, this stable region increases when we decrease the eccentricity of the binary (which is not known for the systems in consideration). During the formation of the giant planet, the gas disk was certainly perturbed to some extent, but taking into account that binary systems with a stellar separation of only 20~AU (e.g., $\gamma$ Cephei; \citealp{Kley08}) can host also a giant planet, we assume that it would have been possible to form a giant planet and Io-sized objects in a disk in both systems, WASP12 and HD189733.

\subsection{Dynamical stability}

\begin{table}
  \caption{Planet \& star parameters for WASP-12b and HD~189733b.}
  \begin{center}
    \leavevmode
 \begin{tabular}{@{}lcc}   
   \hline         
  Exoplanet				& WASP-12b 	& HD~189733b    \\ \hline 
  Orbital distance [au]	&  0.023	&  0.031 	 	  \\
  Planetary mass [$M_{\rm Jup}$]	& 1.404	& 1.138 	 	  \\
  Star mass [$M_{\rm Sun}$]			& 1.35 	& 0.8 	 	  \\ \hline
    \end{tabular}
  \end{center}
   \medskip
   \label{t1}
\end{table}

In this Section, we investigate the stability of possible Trojan satellites in the WASP-12 and HD~189733 systems. First we discuss the main architecture of these two systems.

In this study, we do not consider the influences of WASP-12~BC and HD~189733~B due to their large distances from their primaries and small masses, and refer to WASP-12~A and HD~189733~A as simply WASP-12 and HD~189733, respectively. This assumption seems to be valid because such distant companions do not significantly influence the dynamics of WASP-12b, HD~189733b, and their possible satellites.

\citet{Maciejewski13} speculated that the WASP-12 system contains an additional planet, WASP-12c, with a mass of approximately 0.098~$M_{\rm Jup}$, a semi-major axis of 0.0507~au, and an eccentricity of 0.284. With the presence of a second planet, WASP-12b would have an eccentricity of $e \approx 0.0447$. We note that \citet{Haswell12} also looked for the presence of a second planet, without finding any. Given that the presence of a second planet and of an eccentric orbit for WASP-12b is still speculative, we assume here a zero eccentricity for WASP-12b and no second planet.

We study the stability of the assumed satellites and planetesimals by means of numerical simulations applying the Bulirsch-Stoer integration method with adaptive spacial step-size. The trajectories of bodies are integrated, and the distance between them is calculated in 3D. Depending on the distance between the bodies, a smaller or bigger integration step-size is chosen. As a dynamical model, we use either the elliptic restricted three body problem (ERTBP) or the three body problem (TBP). In the ERTBP the motion of a mass-less body ($m_3$) is studied in the gravitational field of two massive bodies ($m_1$ and $m_2$), i.e. the host-star and the giant planet. Since $m_3$  does not influence the motion of the massive bodies, $m_1$ and $m_2$ move on Keplerian orbits around their center of mass. This model is commonly used in studies of celestial mechanics and gives quite reasonable results if the mass of the third body is small compared to that of the other two. In case the third body has a non-negligible mass, it is advisable to use the TBP model, where gravitational interactions between all three bodies are studied. 

We check the long-term stability of $m_3$ by calculating the Fast Lyapunov Indicator (FLI) of each orbit. The FLI is a well known chaos indicator, which was introduced by \citet{Froeschle97} and allows a fast determination of chaotic motion. 
The application of such a chaos indicator allows a shorter computation time. Froeschl\'{e} and his co-workers studied the behavior of the FLIs in detail and found out that the computation time can be 200 times shorter than for pure orbital computations (Froeschl\'{e}, private communication).
To compute the FLIs one has to calculate the variation equations in addition to the orbit, and the largest tangent vector defines whether an orbit is chaotic or not. This method has already been successfully used to determine the stability of Trojan exoplanets \citep{Dvorak04}. Using the orbital parameters listed in Table~\ref{t1} for the two exoplanetary systems, we determine the stable areas around the L$_4$ and L$_5$ Lagrangian points. The computation time for the FLI calculations is between $10^3 $ and $10^5$ years which corresponds to $3.3 \times 10^5$ to $3.3 \times 10^7$ periods of WASP-12b and to $1.6 \times 10^5$ to $1.6 \times 10^7$ periods of HD~189733b. As we did such computations for several orbits over 500 000 and 1000 000 years (meaning 1~year~=~365.25days) which can be considered as orbital computations over 100 000 000 and 200 000 000 periods of the detected giant planet where nothing changed in the orbital motion compared to the shorter computations, we can assume that the results of the orbital computations are valid over billions of years\footnote{Computations of close-in planets over billions of years are not feasible, as the stepsize would be too small and we risk to accumulate computational errors.}.

In addition, we used another numerical integrator applying the Gauss Radau method with adaptive step-size, which considers relativistic effects. Taking into account these effects, we observe that the stable areas of the small bodies are restricted to the orbital distance of the planet, which means, that small changes in semi-major axis lead to the bodies escaping from the system. The resulting stable areas around $L_4$ and $L_5$ in the two exoplanetary systems are displayed by the blue curves in Figure~\ref{f_2D}. The two panels show the planar case plotted in Cartesian coordinates, with the host-star in the middle and the observed giant planet marked by a black dot. The different size of the objects reflects the parameters listed in Table~\ref{t1}.

\begin{figure*}
\includegraphics[width=2.0\columnwidth]{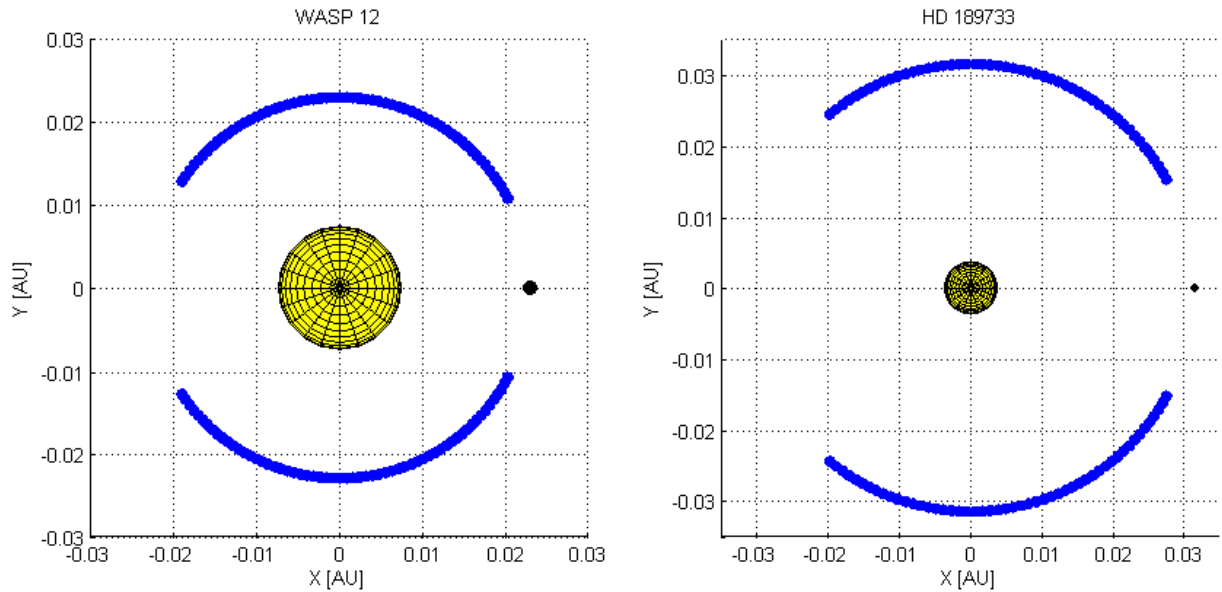}
 \vspace{3.5cm}
 \caption{The blue region represents the stable area for tadpole motion of Io-sized satellites around the L$_4$ and L$_5$ points in the orbital plane of the planet. The yellow circle in the center represents the star, the black point represents the planet (the scaling is according to the system parameters, see Table~\ref{t1}). Left panel: WASP-12b. Right panel: HD~189733b.}
     \label{f_2D}
\end{figure*}

\begin{figure*}
\includegraphics[width=2.0\columnwidth]{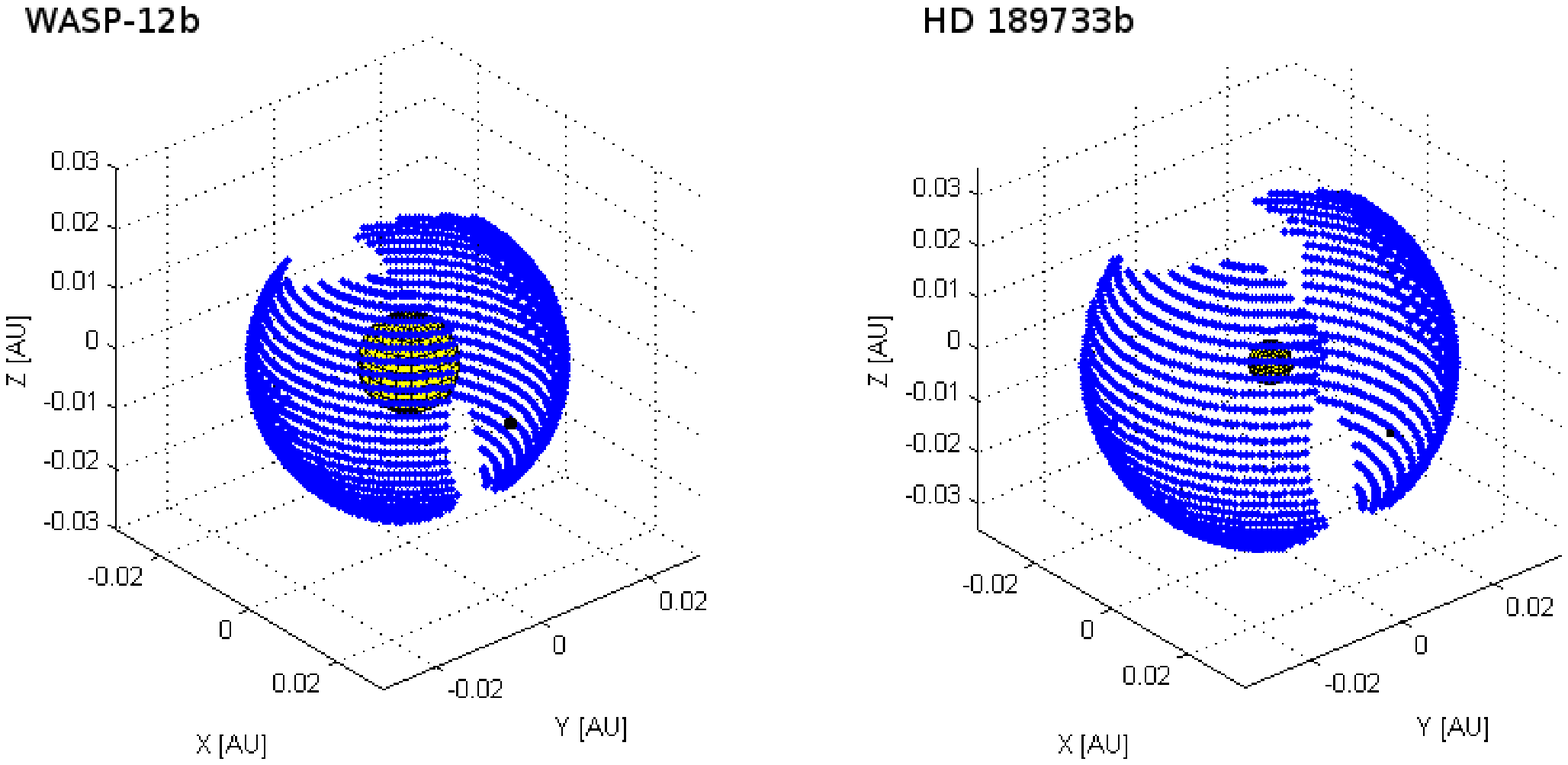}
 \vspace{3.5cm}
 \caption{Same as Figure~\ref{f_2D}, but for inclined orbits of Trojan satellites. Every blue point represents a stable position. One can clearly see that the stable region almost completely covers the stellar disc. Left panel: WASP-12b. Right panel: HD~189733b.}
     \label{f_3D}
\end{figure*}

Even if the stable motion of an Io-sized body is restricted to the orbital distance (i.e., semi-major axis) of the giant planet, our numerical investigation shows that the orbit of $m_3$ could be inclined. The result of this study is summarized in the two 3D plots of Figure~\ref{f_3D}, which show clearly that the stable areas provide nearly complete coverage of the stellar disc. The computation time for this study is $10^6$ years which means that the bodies can be considered long-term stable. Our  numerical simulations have shown, that either the small bodies escape quite quickly or they are stable for the whole computation time. We couldn't observe escape orbits that are stable for some 100 thousands years before they are ejected from the systems. In addition we calculated the FLIs for orbits close to the stability border, where the time evolution of the FLIs did not show any sign of chaos.

In addition to studying a single Io-sized body moving in one of the blue areas shown in Figure~\ref{f_3D}, we investigated also the stability of a Trojan swarm of 100 objects with diameters of 100~km. These objects were distributed uniformly over the entire blue area shown in Figure~\ref{f_3D}. This simulation showed stable motion for almost all objects of the Trojan swarm in the case of circular or nearly circular motion. 

The panels of Figure~\ref{f_inc} give additional information about the elongation of a ``Trojan orbit'' for different initial pericenter distances and inclinations. There is a large area present in both plots, where only small elongations were found for the regions of the allowed orbits (i.e., the narrow blue regions). One can see that highly inclined orbits are bound to a small area around $L_4$ or $L_5$. Orbits with larger elongation are marked by the green, yellow and orange areas. Red indicates unstable orbits with an elongation of $180^\circ$ or larger. One can see that the boundary of the allowed regions is quite constant with a small shift for inclinations between 10 and nearly 40 degrees. We see an increase of the stable zone for inclinations between 10 and 35 degrees where the orbits are quite elongated (orange region). We did not find any stable horse-shoe orbits for the two considered systems, however, one can not in principle exclude satellites with such an ``exotic'' orbital configuration.

One should also note that we did not consider the influence of tides on the orbital stability of Trojan objects. Due to the close proximity of HD~189733b and especially WASP-12b to their host stars, the influence of tides can be significant. At present, there is no theory available for the tidal perturbation induced by two bodies (the star and the giant planet) on the motion of the third one (the satellite) that can be applied here. Therefore, the calculation of this effect is beyond the scope of the present study and may be the subject of a future investigation.

\begin{figure*}
  \begin{center}
		\includegraphics[width=2.0\columnwidth]{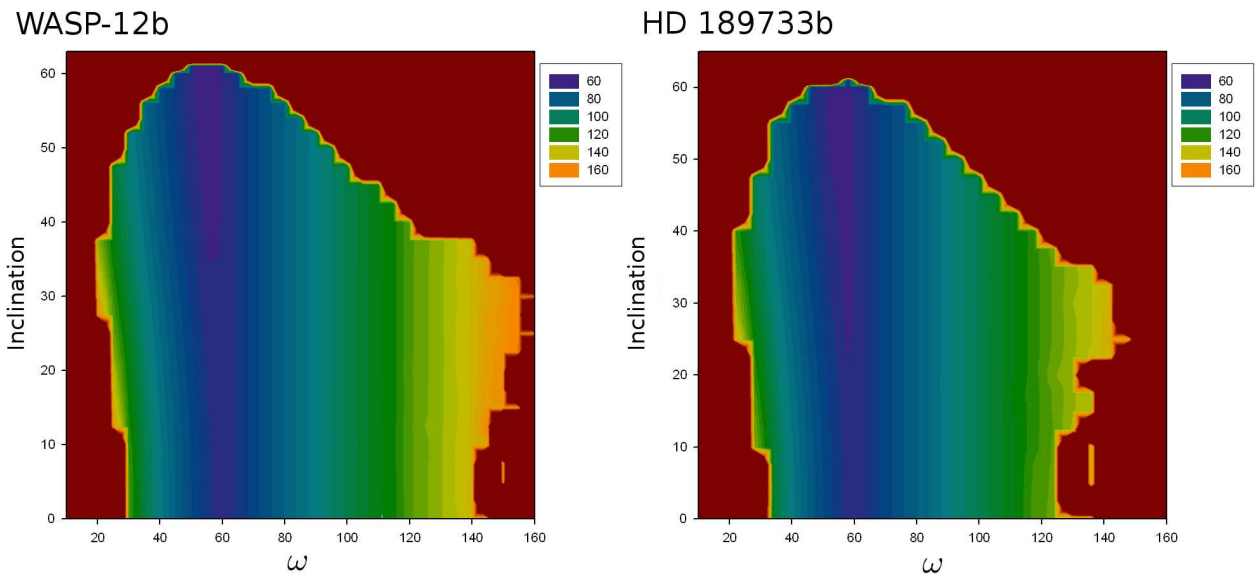}
        \caption{Range of possible inclinations of stable Trojan orbits. The color code corresponds to possible elongations of stable orbital motion given in degrees. Red color marks the unstable area. The y-axis shows the inclinations, the x-axis shows the angle between the satellite and the giant planet, $\omega$. Both axis units are given in degrees. See text for details. Left panel: WASP-12b. Right panel: HD~189733b.}
     \label{f_inc}
  \end{center}
\end{figure*}

\subsection{Photogravitational stability}
\label{subsec_photo}

In hot Jupiter systems, relativistic effects are not the only ones which play an enhanced role for Trojan's stability in comparison to Jovian Trojans. Here we show that while km sized Jovian Trojans are photo-gravitationally stable for its age, even 10--100~m sized bodies can escape the Lagrangian points $L_4$ and $L_5$ of the systems considered here within several million years. This result contradicts the conclusion made by \citet{Moldovan10} that the influence of the Yarkovsky effect -- a delayed re-emission of the incoming sunlight -- is negligible for all bodies with the size exceeding 1~m. 

It is a well known fact that stellar radiation can alter the dynamics of orbiting bodies. Not all objects are equally affected by stellar radiation. For dwarf planets and planets in the Solar System, the effect of solar radiation pressure (SRP)  is practically negligible. The key quantities that determine whether SRP plays a role in the dynamical behavior of the system are 
\begin{enumerate} 
\item the effective area-to-mass ratio $\alpha=S/m$ of the object under consideration, e.g. the Trojan asteroid (Solar Sails have a large area-to-mass ratio whereas large spheres have generally small $\alpha$ values), 
\item the amount of effective insolation the object receives (luminosity $L$ and distance to the radiation source $d$), 
\item and the relative importance of the gravitational acceleration compared to the SRP.
\end{enumerate}

\citet{Schuerman80} has introduced the efficiency parameter $\beta$ comparing the relative importance of the acceleration that can be expected from the SRP to the gravitational acceleration in the system
\begin{equation}
\beta\approx \frac{\alpha \;L }{4 \pi\; c\; G\; M}.
\end{equation}
Here, $c$ is the speed of light in vacuum, $G$ the standard gravitational constant, and $M$ the mass of the host star.
In contrast to \citet{Schuerman80} we have simply assumed that all the spectral energy is absorbed and re-emitted by the Trojan, not just a specific spectral range, see e.g. \citet{Veras15}. If $\beta > 1$, then SRP becomes more important than gravity in the system. However, even small values of $\beta$ can have drastic effects on the  system's dynamics.

As above, we model the dynamics in considered systems using the circular restricted three body problem (CR3BP), i.e. the star and the planet are much more
massive than the Trojan asteroid, the orbit of the planet is roughly circular and co-planar with the Trojan's. Both HD~189733b and WASP-12b have a very low eccentricity and the M-dwarf companions are several tens of AU away from the system. Furthermore, we require that the mass of the star be much larger than the mass of the planet (hot Jupiter), condition which is satisfied in both cases.

For this setup, \citet{Schuerman80} showed that SRP and Poynting-Robertson Drag cause the Lagrangian equilibrium points $L_4$ and $L_5$ to be displaced from their original position, and  to be unconditionally unstable. However, the time it takes for small perturbations to grow can be very long. \citet{Schuerman80} derived the following expression for the so-called $e$-folding time $\tau$, i.e. the time it takes for a small perturbation (out of the equilibrium position) to grow by a factor of $e\approx 2.718$.

\begin{equation}
\tau=\frac{(1-\beta)^{2/3}\; c\; d^2}{3\; \beta\; G\; M}.
\end{equation}

Here, $\tau$ is in seconds. For our Trojans we assume a spherical shape as well as a low thermal inertia and slow rotation. This causes the Yarkovsky effect and SRP to align directions and add up to a mere factor $Q$ between 1 and 2 in the effective radiation cross section \citep[equation (30) in][]{Veras15}. Therefore, the area-to-mass ratio of our Trojans becomes
\begin{equation}
\alpha=S/m=\frac{Q \pi r^2}{4/3\pi r^3 \rho} = \frac{3 Q }{4 r\rho},
\end{equation}
where $\rho$ is the bulk density of the Trojan, which we assume to be $\rho=2700$~kg/m$^3$. Spherical bodies have a very small radiative cross section-to-mass ratio. In this respect, our results are an optimistic limit regarding the resident times in the vicinity of the Lagrangian points. Given the same density, any other particle shape should lead to a stronger reaction to the stellar radiation pressure and, thus, a faster escape. Figure~\ref{f_yark} shows the resulting e-folding times for Jupiter Trojans in the Solar System and possible Trojans in the WASP-12 and HD~189733 systems.
It is clear that small exo-Trojan particles in the systems under consideration leave the vicinity of the Lagrangian points extremely fast. Choosing $Q=1$ or $Q=2$ is irrelevant given the fact that the $e$-folding times for the exo-systems are orders of magnitude shorter (we have chosen $Q=2$).

Since the hot Jupiters are very close in both exo-systems, the SRP is strong enough to fling out meter sized asteroids on timescales of several million years. Keeping small asteroids \textit{in situ} for the age of the respective systems is, thus, difficult. Trojans beyond kilometer size that have formed very rapidly or been captured during the migration phase could potentially survive long enough to still be around. For small bodies, radiation pressure reduces the size of the stability region \citep{Markellos00}.

\begin{figure}
  \begin{center}
		\includegraphics[width=0.7\columnwidth, angle=-90]{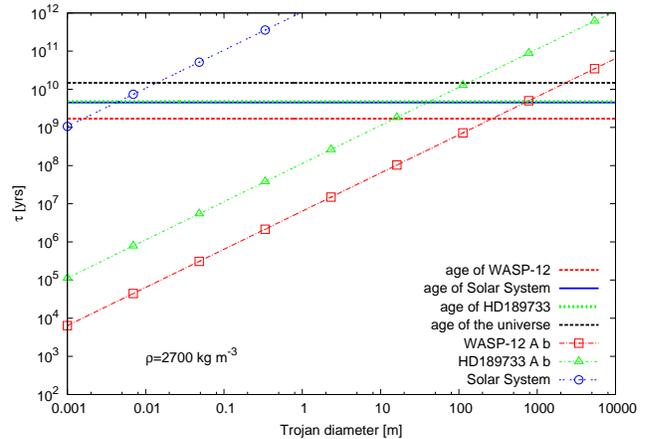}
        \caption{e-folding times ($\tau$) for Trojans of different sizes. Jupiter Trojans in the Solar System are represented by circles, possible Trojans in the WASP-12 and HD~189733 systems are denoted by squares and triangles, respectively. Jupiter Trojans with radii larger than 10 meters can remain in the vicinity of the Lagrangian points $L_4$ and $L_5$ for the age of the Solar System. In contrast, even km sized asteroids have relatively short e-folding times in the investigated exo-systems. In other words, small Trojans in WASP-12 and HD~189733 systems are dynamically unstable on relatively short timescales. Note that the age of HD~189733 is very close to the Solar system age.}
     \label{f_yark}
  \end{center}
\end{figure}

Therefore, we conclude that the presence of small bodies in the WASP-12 and HD~189733 systems can be compromised by photogravitational effects. Although the existence of many middle-sized objects ($\ge$10~km) is not excluded, we focus on two Io-sized satellites orbiting in L$_4$ and L$_5$. 

The half-surface of an Io-mass satellite, which is exposed to the stellar radiation (assuming that the satellite is tidally locked and only one side is irradiated), is calculated as $S_{\rm sat} = 2 \pi R_{\rm sat}^2$, assuming a density of a rocky undifferentiated body (2700~kg~m$^{-3}$). This gives an estimate of $S_{\rm sat}  \approx 1.5 \times 10^{12}$~m$^{2}$ or $A_{\rm sat} \approx 3 \times 10^{12}$~m$^{2}$, if we assume that both $L_4$ and $L_5$ host a Trojan satellite.

For two Trojan swarms of 100 objects with the radius of 100~km (200 in L$_4$ and L$_5$ together), one obtains an estimate of $S_{\rm swarm} \approx 1.25 \times 10^{13}$~m$^{2}$. However, we consider two Io-sized objects to have a better dynamical long-term stability.

\section{Volatile outgassing of co-orbital satellites}
\label{sec_outgassing}

In the model proposed by \citet{Ben-Jaffel14} the outgassing rates of $\approx$10$^{28}$ Mg{\sc ii} ions s$^{-1}$ and $\approx$10$^{29}$ C{\sc ii} ions s$^{-1}$ were assumed to reproduce the early ingress observed in the UV with HST for the transits of WASP-12b and HD 189733b. Here we describe the volcanic outgassing of Io and investigate if this mechanism can operate for WASP-12 and HD~189733 systems.

\subsection{Io's volcanic activity}
\label{subsec_Io}
In the Solar system, Io, the innermost of the four large satellites of Jupiter, is the most volcanically active body (e.g., \citealp{McEwen98}). The tidal heating generated due to Io's orbital eccentricity produces at its surface hundreds of volcanic spots and extensive lava flows. Three main mechanisms for generating an atmosphere on Io are proposed: frost sublimation, surface sputtering, and active volcanism (e.g., \citealp{Chrisey88,Wong00,Bagenal11}). During a major eruption, lava flows tens or even hundreds of kilometers long can be produced, which consist mainly of basalt silicate lavas with either mafic or ultramafic Mg-rich compositions. As a by-product of this activity, sulfur, SO$_2$ and silicate pyroclastic material are ejected up to 200~km into space at a rate of $\approx$9$\times$10$^{28}$ s$^{-1}$, providing material for interaction with the Jovian magnetosphere. SO$_2$ molecules are dissociated and important fractions are lost in the form of neutral O and S atoms. These materials form extended clouds along Io's orbit, and over time are ionized by two main processes: electron impact and charge exchange reactions. The ionized components are rapidly accelerated under the effect of the Jovian magnetic field and form a ring-like structure known as Io plasma torus, where electron density reaches $10^4$~cm$^{-3}$ \citep{Boudjada14}. To supply the electron densities observed in the plasma torus, Io ejects mainly S$^+$ and O$^+$ ions at rates of $\approx$2$\times$10$^{28}$ s$^{-1}$ \citep{Thomas04}. 

Besides SO$_2$, millimeter wavelength observations indicate also the presence of Na-, Cl-, and K-bearing volatiles \citep{Schaefer05}. Other elements related to chondritic abundances such as Li, Br, Rb are negligible compared to the above mentioned species \citep{Fegley00,Moses02,Schaefer05}. 

The observed S$^+$ and O$^+$ Io ejection rates are of the same order as the outgassing rates of Mg{\sc ii} and C{\sc ii} necessary to explain the UV early ingress observations of WASP-12b or HD 189733b assuming the exomoon cloud hypothesis \citep{Ben-Jaffel14}. However, the observed elements are different. On Io and its volcanic environment, Mg and also C are not observed. Only the observational composition of Jovian dust stream particles reveals a weak evidence for mass lines of metals that typically accompany rocky minerals such as Mg, Fe, Ca, Al and Ni \citep{Postberg06}. Therefore, we conclude that the volcanic activity is probably not the most plausible mechanism for the production of the amount of Mg and C required to explain the observations.

\subsection{Outgassing from molten lava surfaces: a source of Mg and Ca}
\label{subsec_lava}

Under hot Jupiters conditions, besides of volcanic activity and sputtering, one can argue that the surface of a satellite that points towards the host star consists of lava similarly to what expected for extreme hot rocky exoplanets such as CoRoT-7b (e.g., \citealp{Guenther11,Mura11}). Depending on the crust composition, at the temperature of WASP-12b elements such as Mg, Ca, Si, Na, K and others can be thermally released in the surrounding environment where they are ionized by stellar radiation or electron impacts within seconds (Ca), hundreds (Mg) or thousands of seconds (Mg$^+$, Ca$^+$), trapped in the interplanetary magnetic field and dragged away from the system \citep{Mura11}. Mg and Ca are believed to form an ionized tail behind CoRoT-7b \citep{Mura11}. \citet{Miguel11} investigated the dominant species in the atmospheres of hot rocky planets depending on their temperature. According to their results, the gaseous envelope surrounding HD~189733b Trojans should be very sparse because of the comparably low equilibrium temperature ($\approx$1419~K, \citealp{Crouzet14}) leading to low outgassing rates. It should be dominated by such elements as monatomic Na, O$_2$, monatomic O, and monatomic Fe. Rocky Trojans orbiting WASP-12b (equilibrium temperature of about 2578~K, \citealp{Sing13}) should possess an atmosphere of a different type, dominated by such gases as monatomic Na, O$_2$, monatomic O, SiO, monatomic Mg, and monatomic Fe. This is in agreement with a new study of \citet{Ito15}, according to which Na, K, Fe, Si, SiO, O, and O$_2$ are the major atmospheric species outgassed from molten surfaces of rocky Super-Earths.

If we assume a Mercury composition for Trojan objects and the Mg escape rate of $\approx 10^{30}$~s$^{-1}$ from the CoRoT-7b mineralogical exosphere study of \citet{Mura11}, we obtain an escape rate from the atmosphere per square meter of $\approx 7.5 \times10^{14}$~s$^{-1}$~m$^{-2}$. The study of \citet{Mura11} is based on the results obtained by \citet{Schaefer09}. \citet{Schaefer09} used a specialised MAGMA code to compute the pressure determined by gas-melt equilibrium and applied it to CoRoT-7b magma ocean at different temperatures. They assumed that the planet is a terrestrial-type planet rather than the core of a giant planet, and has an Earth-like composition. According to their results, Mg is an abundant element in the mineral atmosphere, especially if one assumes the atmosphere being constantly lost. The same code was used by \citet{Miguel11} to calculate the amount of atmosphere outgassed from magma oceans at the surface of rocky planets in a wide mass range. Assuming Io's gravity, the Mg column density for the Trojans at 2578~K is in the range of $3-4.8 \times 10^{23}$~cm$^{-2}$ depending on the crust composition. At 1419~K (HD~189733b Trojans), the Mg column density is in the range of $2.6-3.6 \times 10^{13}$~cm$^{-2}$, i.e., many orders of magnitude lower. For Ca at 2578~K and 1419~K, the column densities are $3.7-3.8 \times 10^{18}$~cm$^{-2}$, and $8.0 - 8.7 \times 10^{7}$~cm$^{-2}$, respectively (all column densities are provided by Yamila Miguel, private communication). These column densities don't differ much from the ones shown in Fig. 3 of \citet{Miguel11} for a planet ten times more massive than the Earth, meaning that the Io-sized Trojans can also outgas the material quite effectively and probably outgas the same amount of atmosphere from a surface unit. To estimate the amount of material lost to space, one should calculate the atmospheric escape similar to the study of \citet{Mura11}. The escape is probably more significant for Trojan satellites possessing only weak gravity in comparison to the heavy planet CoRoT-7b studied by \citet{Mura11}. However, proper escape calculations are beyond the scope of the present study and will be performed in the future. Here, we only scale the escape rates for CoRoT-7b from \citet{Mura11} to the total surface of two Io-sized bodies assuming that they outgas the same amount of atmosphere per surface unit as we have shown above. This can probably underestimate the escape rate, because the material can easier escape a body with a weak gravity, i.e., the Trojans. Being scaled to a total surface of two Trojan satellites of $S_{\rm sat} \approx 3 \times 10^{12}$~m$^{2}$ estimated above, the total Mg outgassing rate is of $\approx 2.2 \times 10^{27}$ s$^{-1}$. For two Trojan swarms with the total surface of $S_{\rm swarm} \approx 1.25 \times 10^{13}$~m$^{2}$, it gives a total Mg outgassing rate of $\approx 9.3 \times 10^{27}$ s$^{-1}$. According to \citet{Mura11}, the outgassing rate of Ca is at least 10 times lower, yielding production rates of $\approx 7.5 \times10^{13}$~s$^{-1}$~m$^{-2}$ and the total outgassing rate of $\approx 2.2 \times 10^{26}$ s$^{-1}$ and $\approx 9.3 \times 10^{26}$ s$^{-1}$ for two Trojan satellites and two Trojan swarms, respectively. According to the estimate of \citet{Miguel11}, the amount of the outgassed Ca atoms may be lower. We also note that the outgassing rates for the asteroid swarm may be even less precise than the ones for two Io-sized Trojans. Mg is ionized within a few hundreds of seconds, which provides a sufficient source of Mg{\sc ii} ions. The photoionization time of Ca atoms is only a few seconds. Fig.~6 and 8 in \citet{Mura11} illustrate the appearance and shape of the simulated Mg+ and Ca+ tails in CoRoT-7b system. These ions can be ionized on a short time scale to produce Mg{\sc ii} and Ca{\sc ii} ions. For Trojans in WASP-12 system, the tails of Trojan satellites can look qualitatively the same, but the lower outgassing rates and a slightly different stellar wind direction have to be accounted for.



The destabilizing of small Trojan asteroids by photogravitational effects and other mechanisms require a presence of bigger satellites, although additional numerical simulations are required to model the distribution of the outgassed material along the orbit of WASP-12b and the coverage of the stellar disc. The presence of many Trojans with the size of a few hundred km is not excluded. High outgassing rate, temperature, and orbital speed of the satellites would lead to a distribution of ions over a wide spacial range. These simulations will be a focus of a future study.

In the case of the WASP-12 system, \citet{Haswell12} estimated the outgassing rate necessary to completely absorb the stellar emission in the Mg{\sc ii}\,h\&k resonance lines obtaining  $\dot{m} \ge 10^{37}$~s$^{-1}$, which is a few orders of magnitude higher than the one estimated assuming the evaporation from the total surface of both satellites and an asteroid swarm\footnote{Note that there is an algebric error in the calculation of \citet{Haswell12}, which has been corrected for this estimation.}. On the other hand, the estimation of \citet{Haswell12} is highly uncertain and a difference up to a few orders of magnitude can be expected.

Such a high number can not be provided by outgassing rates of Trojan satellites. The planet itself can also not provide a magnesium source of the required power. If one applies the hydrodynamic code describing the upper atmosphere and calculating the escape \citep{Erkaev13} of WASP-12b, one obtains a loss rate of $1.5 \times 10^{35}$~particles (hydrogen atoms) per second for the predominantly hydrogen atmosphere. This number is in a very good agreement with other studies for mass-loss rates for close hot Jupiters, e.g., \citet{Shaikhislamov14}. If one assumes a Mg solar abundance mix of $10^{-4}$, this gives an upper limit of $1.5 \times 10^{31}$ magnesium atoms per second. Besides, one should note that not every atom of magnesium is dragged up by the escaping hydrogen, leading probably to additional several orders of magnitude decrease of the maximum amount of magnesium atoms possibly provided by the planet. On the other hand, \citet{Ben-Jaffel14} were able to reproduce the observations with the outgassing rate of $\approx 10^{28}$ Mg{\sc ii} ions per second, which illustrates that the estimates of the required amount of ionized magnesium atoms are highly uncertain and differ by many orders of magnitude.

One should note that if Mg is released in such amount by Trojan satellites, then other silicate-type elements like Na, O, Ca and Si should also be observed in large quantities \citep{Schaefer09}. This is in good agreement with the complete absorption of the stellar Mg{\sc ii}\,h\&k and Ca{\sc ii}\,H\&K line core emission in WASP-12 \citep{Haswell12,Fossati13}.

\subsection{Outgassing of carbon in HD~189733 system}
\label{subsec_carbon}

It is quite difficult to provide the necessary high outgassing rates of C atoms to explain the early ingress observed for HD~189733b. None of the studies considering mineral atmospheres of rocky bodies predicts C to be present in large amount \citep{Mura11,Miguel11,Ito15}. \citet{Schaefer09} investigated the composition of early close-in rocky cores by chemical equilibrium and chemical kinetic calculations. They modeled the chemistry of the volatiles released by heating different types of carbonaceous, ordinary and enstatite chondritic material as a function of temperature and pressure, finding that volatile elements such as H, C, N, S, and Cl should be lost from the young planetary body, so that only silicate atmospheres remain. This is in agreement with a recent study of \citet{Lammer14} who investigated the origin and stability of magma ocean related outgassed exomoon atmospheres around gas giants in the habitable zone of their host stars. It was found that even at orbital locations of 1 au, even larger Mars-size exomoons would lose their outgassed volatiles and dissociation products such as H, O and C atoms related to partial surface pressures of $\approx$30-150 bar within a few million years. 

Therefore, Trojan satellites seem to be able to provide enough outgassed material (e.g., Mg and Ca atoms) to form a cloud of gas surrounding WASP-12. For the HD~189733 system, this mechanism encounters difficulties due to the lower temperatures. Also, even though the Trojan satellites could be stable in this system, their gaseous envelopes are possibly dominated by constituents other than carbon. For these reasons, we surmise that neither a volcanic moon nor outgassing Trojans can contribute to the early ingress of HD~189733b observed in the FUV C{\sc ii} line. Nevertheless, a possible thin Trojan produced plasma cloud dominated by Na, O, Fe and their ions may leave spectral signatures at other spectral lines.

We also note that surface sputtering can contribute to the plasma production as well, especially for the cooler Trojans in the HD~189733 system. However, the investigation of the contribution of this process to outgassing is beyond the scope of this study.

In the next Section, we will consider the WASP-12b observations in detail.

\section{Plasma cloud surrounding WASP-12}
\label{sec_torus}

WASP-12b is an extreme hot Jupiter for which NUV HST observations showed that the planet atmosphere extends beyond the planet's Roche lobe \citep{Fossati10,Haswell12}. The transit lightcurve showed also the presence of a time-variable early ingress, with egresses compatible with the transit ephemeris \citep{Fossati10,Haswell12}. Given the supersonic motion of the planet inside the stellar wind, \citet{Vidotto10} suggested that the early ingress is caused by the presence of an optically thick bow-shock ahead of the planet \citep[see also][]{Llama2011} sustained either by a planetary magnetic field \citep{Vidotto10} or by the intrinsic expansion of the planet evaporating atmosphere \citep{Bisikalo13}. The HST data also revealed the complete lack of any emission in the core of the Mg{\sc ii}\,h\&k resonance lines, ubiquitous among late-type stars. Note that the anomaly is always present, regardless of the planet's orbital phase. \citet{Haswell12} suggested that this may be caused by the presence of a translucent circumstellar cloud of planetary evaporated material, which absorbs the stellar line emission that shall be indeed present. On the basis of optical data and from measurements of the interstellar absorption along the WASP-12 line of sight, \citet{Fossati13} confirmed \citet{Haswell12}'s suggestion and showed that also the core of the Ca{\sc ii}\,H\&K resonance lines of WASP-12 presents the same anomaly observed for the Mg{\sc ii}\,h\&k lines.

In order to be able to effectively absorb the stellar Mg{\sc ii}\,h\&k and Ca{\sc ii}\,H\&K line core emission, the circumstellar cloud should cover most of the observed stellar disk: the cloud would need to extend for about one stellar radius above and below the planet orbital plane. The very extended asymmetric planet atmosphere (\citealp{Bisikalo13} showed that the planet atmospheric gas may be present as far as 0.8 stellar radii away from the planet), the short planet orbital period of about $\approx$1\,day (replenishing therefore the cloud at every orbital period of freshly evaporated planetary material, which would otherwise leave the system under the action of stellar wind and radiation pressure), and the expected high temperature of the cloud (avoiding the cloud to collapse into a disk; \citealp{B2002}) suggests that this may be in principle possible, but dedicated MHD simulations need to be carried out in order to test this hypothesis.

As we show below, both Mg and Ca would evaporate from the molten surface of the rocky bodies present in the system. In this study, we focus on objects located at the $L_4$ and $L_5$ Lagrangian points. The outgassed material would be quickly ionised by the stellar UV radiation \citep{Mura11}, possibly giving rise to the absorption signatures observed in the core of the Mg{\sc ii}\,h\&k and Ca{\sc ii}\,H\&K resonance lines (Fig.~\ref{f_haswell}). The presence of evaporating Trojan asteroids or satellites orbiting around the star along with the planet may easilly lead to the formation of a stable cloud, thick enough to cover the entire stellar disk. As shown below in Section~\ref{sec_stability} (see Figures~\ref{f_2D} and \ref{f_3D}), Trojans would lie around the $L_4$ and $L_5$ Lagrangian points, and extend for several planetary radii in the direction perpendicular to the planet orbital plane. As a result, at each point on the circumstellar cloud, at most only half a planet orbital period ($\approx$0.5\,days) would be needed before newly evaporated material replenishes the cloud. In addition, the fact that the Trojan swarm would also stably lie at large distances from the planet orbital plane would naturally lead to the formation of a cloud thick enough to cover most of the stellar disk. In the case of only two Trojan satellites, it would be less likely for the outgassed material to arrive at such lage distances from the orbital plane, but detailed modelling is needed to completely exclude it.

Due to the small masses and, therefore, weak gravity of Trojan satellites, newly outgassed atoms can easily escape to space, where they are ionized by intense stellar radiation and electron impacts. The stable regions around L$_4$ and L$_5$ in the WASP-12b and HD~189733b systems are very extended, so that a satellite can approach quite close to the planet. Because the satellites move around the L$_4$ and L$_5$ in more or less elongated orbits depending on the initial distance to the Lagrange point, their outgassed plasma could be distributed along the orbit and range of inclination angles. The orbital period around L$_4$ and L$_5$ is of between 14 and 16 days for the two systems considered here. This movement would cause the absorption observed in the Mg{\sc ii}\,h\&k and Ca{\sc ii}\,H\&K line cores of WASP-12 and could explain why the absorption does not depend on the planet's orbital phase. However, this mechanism requires high temperatures that are able to melt the surfaces of rocky Trojans. We show that this mechanism can operate for the WASP-12 system, but it is only marginally possible for HD~189733.

\citet{Lanza2014} used the presence of circumstellar condensations of planetary evaporated material to explain the correlation between stellar activity and planet surface gravity for systems hosting hot Jupiters \citep{Hartman2010,Figueira2014,Fossati15}. Condensations formed by Trojan asteroids or satellites evaporated material, rather than planetary, would only partly conflict with the model proposed by \citet{Lanza2014}. The evaporation of rocky material requires very high temperatures, achievable only for the most close-in planets orbiting around the hottest stars, which are only a minority in the sample of systems considered by \citet{Hartman2010}, \citet{Figueira2014}, \citet{Lanza2014}, and \citet{Fossati15}. In the context of the model proposed by \citet{Lanza2014}, planetary evaporated material would form the condensations and the evaporating Trojans would just increase their absorbing power for the systems hosting extremely close-in planets, such as WASP-12.

Evaporating Trojan asteroids or satellites would also affect the bow-shock present ahead of the planet, suggested by \citet{Vidotto10} to explain the early-ingress phenomenon. On the side of the planet, the bow-shock region would be filled with planetary and stellar wind particles. In addition, along its motion around the star, the bow-shock would continuously encounter rocky evaporated material, which would accumulate for a certain time on the external side of the bow-shock region. This will increase the density and lower the temperature in the bow-shock region and hence the probability of the bow-shock to absorb the stellar radiation and cause an observable early-ingress.


\section{Exomoons as an outgassing source: hill radii and Roche limits of exoplanets}
\label{sec_Hill}

In this section, we consider whether classical exomoons orbiting WASP-12b and HD189733b may exis. We also check if the orbital configurations proposed by \citet{Ben-Jaffel14} can be stable. All moons have to orbit inside the Hill radius of their planets to be gravitationally bound to them. Moreover, it has been shown that even the limit of the Hill radius is too optimistic (e.g., \citealp{Domingos06,Donnison14}, and references therein). 

If the ratio of the planet mass to the star mass, $q$, is $q \ll 1$, one can estimate the Hill radius, $r_H$, of an exoplanet in units of orbital separations as
\begin{equation}
	r_H \approx \sqrt[3]{\frac{q}{3}},~q \ll 1.
	\label{e_rH}
\end{equation}
The maximum possible separation between an exoplanet and a satellite in units of Hill radii can be written as
\begin{equation}
	a_p \approx 0.4895 (1.0000 - 1.0305 e_p - 0.2738 e_s)
	\label{e_amax_prog}
\end{equation}
for prograde satellites and
\begin{equation}
	a_r \approx  0.9309 (1.0000 - 1.0764 e_p - 0.9812 e_s + 0.9446 e_p e_s)
	\label{e_amax_retro}
\end{equation}
for retrograde satellites \citep{Domingos06}. Here, $e_p$ and $e_s$ are the eccentricities of the planet and the satellite, respectively. Little is known about the possible eccentricities of exomoons, but it is known that close-in giant planets often have zero or very small eccentricities because of tidal dissipation (e.g., \citealp{Naoz11}). Therefore, for putative hot Jupiter moons one can assume $a_p \approx 0.49~r_H$ and $a_r \approx 0.93~r_H$. In general, eccentric and/or inclined satellites are less stable \citep{Donnison14}. 

\begin{table}
  \caption{Hill radii, maximum satellite orbital separations, and maximum satellite masses for WASP-12b and HD~189733b.}
  \begin{center}
    \leavevmode
 \begin{tabular}{@{}lcc}   
   \hline         
  Exoplanet				& WASP-12b 	& HD~189733b    \\ \hline 
  $q = M_{p}/M_{s}$	& $1.0 \times 10^{-3}$	 & $1.4 \times 10^{-3}$   \\
  Star age [Gyr]			& 1.7$\pm$0.8 	& 4.75$\pm$0.75 	  \\
  Hill radius $R_{Hill}$ [$R_{\rm p}$]	&  1.96	& 4.6		\\   
  $a_p$ [$R_{\rm p}$]	&  	0.91	& 	2.22	   \\	   
  $a_r$ [$R_{\rm p}$]	&  	1.73	& 	4.23	   \\   
  $R_{Roche}$ [$R_{\rm p}$]	&  1.2	& 1.7		\\   
  $m_{max, p}$ [kg]	&  	0.0	 & $4.6 \times 10^{19}$	    \\ 
  $r_{max, p}$ [km]	&  	0.0	 & 	159 	   \\ 
  $m_{max, r}$ [kg]	&  	$8.35 \times 10^{19}$	& $3.0 \times 10^{21}$	    \\ 
  $r_{max, r}$ [km]	&  	194	 &  640    \\ \hline
    \end{tabular}
  \end{center}
   \medskip
   Here, $q = M_{p}/M_{s}$ is the exoplanet-to-host star mass ratio. Maximum orbital separations, $a_p$ and $a_r$, and Roche limits, $R_{Roche}$, are given in units of planetary radii. The values of $m_{max}$ and $r_{max}$ determine maximum possible masses and radii of the exomoons; the letters "p" and "r" denote the prograde and retrograde satellites, respectively. The maximum possible radii of the moons are calculated assuming the density of a rocky body of 2700~kg~m$^{-3}$. 
   \label{t2}
\end{table}

\begin{figure*}
\includegraphics[width=2.0\columnwidth]{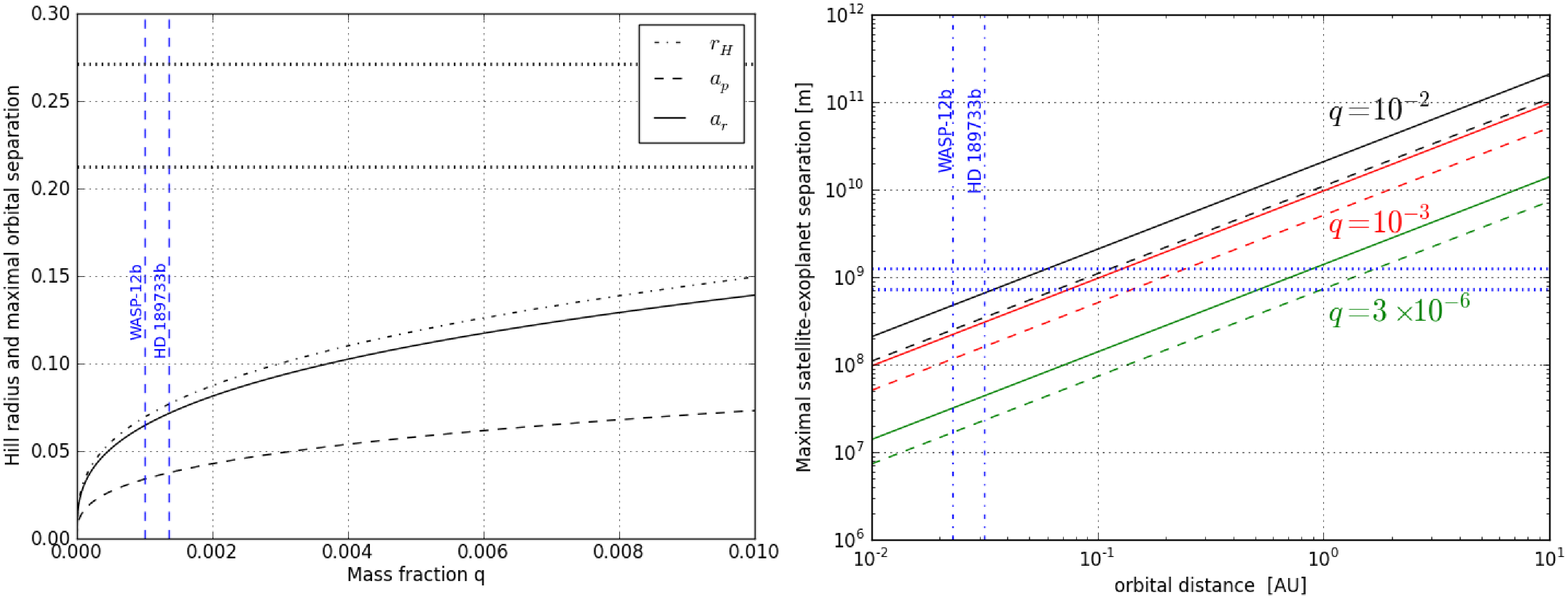}
 \vspace{3.5cm}
 \caption{Left panel: Hill radius $r_H$ (dash-dotted line), maximum possible separation of prograde satellite $a_p$ (dashed line) and retrograde satellite $a_r$ (solid line) depending on the mass fraction, $q$, given in units of planet-star orbital separation. The blue dotted lines mark the location of HD~189733b and WASP-12b. The upper and lower black dotted lines mark the orbital separations assumed by \citet{Ben-Jaffel14} for HD~198733b and WASP-12b satellites, respectively. Right panel: $a_r$ (solid lines) and $a_p$ (dotted lines) depending on the orbital separation for three possible values of mass ratio $q=10^{-2}; 10^{-3}; 3\times10^{-6}$ (black, red, green lines, respectively). The upper and lower dotted blue lines mark the distance of $1.2\times10^9$~m$=16 R_{\rm HD189733b}$ and $7.2\times10^8$~m$=6 R_{\rm WASP-12b}$, respectively.}
\label{f1}
\end{figure*}

Even in stable regions, in each particular case one has to estimate the maximum possible lifetime of a moon with a given mass, which is mostly much shorter than the system's lifetime. If radiative effects are negligible, less massive satellites can survive longer \citep{Barnes02,Weidner10,Domingos06}. Since the orbits of more massive moons decay faster as a result of tidal dissipation of angular momentum, one can estimate the maximum mass a moon can have to survive during the system's lifetime \citep{Domingos06}. The maximum possible mass of a satellite for a given system age, $t$, can be calculated as

\begin{equation}
m_{max} = \frac{2}{13} a_{p,r}^{13/2} \frac{Q_p}{3 k_{2P} t R_p^5} \sqrt{\frac{M_p}{G}}.
\end{equation}

Here, $Q_p$ is the dimensionless tidal dissipation factor, $k_{2P}$ is the tidal Love number, $a_{p,r}$ is the maximum possible prograde or retrograde orbit, $R_p$ is the planetary radius, and $M_p$ is the planetary mass. The value of $Q_p$ is very poorly constrained even for Solar system planets and even more uncertain for exoplanets. Following \citet{Barnes02}, we chose $Q_p = 10^5$ and $k_{2P} = 0.51$, which is consistent with estimate obtained for Jupiter. Q$_p^{-1}$ equals the fraction of tidal energy dissipated during each tidal cycle. For exoplanets, the value of $Q_p$ may differ substantially from the Jovian value in case their interiors differ from Jupiter’s. However, since the mechanism of tidal dissipation and also the interior structure of exoplanets are not yet well understood yet and the other parameters are not available, we assume a value of $Q$ of $10^5$ for our estimate.

Table \ref{t2} summarizes the maximum satellite mass and size, Hill radius and maximum possible separation for prograde ($a_p$) and retrograde satellites ($a_r$) for WASP-12b and HD~189733b. For WASP-12b, the Hill limit extends only until approximately $2 R_p$. Since it is known that this hot Jupiter is very inflated and experiences Roche lobe overflow \citep{Fossati10,Haswell12,Bisikalo15}, it seems very unlikely that this planet can possess a stable satellite even on a retrograde orbit. However, HD~189733b could posses a retrograde moon with a size of approximately 500~km. However, as we show below, even if present, this moon can not provide carbon atoms. As the system ages, the maximum possible size of a satellite decreases \citep{Weidner10}. The ages listed in Table~\ref{t2} are taken from \citet{Chan11} (WASP-12) and \citet{Poppenhaeger13} (HD~189733). For the calculations, we assumed that the average densities of the moons are that of the Earth.

Exomoons can suffer also from formation difficulties. For example, \citet{Namouni10} and \citet{Spalding16} have shown that exomoons formed \textit{in situ} do not survive migration process to close-in orbits. The author concluded, that if any exomoons will be discovered in these systems, they were with a high probability captured from the protoplanetary disk on retrograde orbits around the planets and doesn't consider the existence of prograde moons as probable. This result puts an additional restriction on the exomoon hypothesis.

The left panel of Figure~\ref{f1} illustrates the dependence of the Hill radius and maximum possible orbital separation on the mass ratio between the components. The right panel of Figure~\ref{f1} shows the dependence of $a_p$ and $a_r$ on the exoplanet orbital distance for three values of the mass ratio, $10^{-2}$, $10^{-3}$, and $3\times10^{-6}$. The value $q=10^{-2}$ could only be the case for a very massive planet orbiting a very low-mass star, or a brown dwarf orbiting a solar-type star, but it is shown for comparison. The value of $q=10^{-3}$ is a typical mass ratio for a Jovian planet orbiting a solar-type star, while the value of $3\times10^{-6}$ corresponds to the ratio between the masses of the Earth and the Sun and can be assumed for terrestrial exoplanets. Both $a_p$ and $a_r$ are calculated assuming zero eccentricities of the planet and satellite. WASP-12b and HD~189733b have a value of $q$ close to $10^{-3}$ (Table~\ref{t1}). Note that both orbits assumed by \citet{Ben-Jaffel14} for the possible satellites of WASP-12b and HD~189733b are located well beyond the stability limit. Therefore, the orbital configurations suggested by \citet{Ben-Jaffel14} are not physically possible.

In general, classical exomoons orbiting hot Jupiters can be an outgassing source similar to the Trojan satellite. Considering hot Jupiters, one should account for several factors restricting the sizes and orbital locations of these bodies. The upper limit to the mass of a satellite is put by the tidal interaction and depends on the mass, size and tidal dissipation factor of a planet as well as the age of the system. As the system ages, the upper limit decreases. The lower limit on the size (and also mass which can be inferred from the density) is determined by photogravitational effects (see discussion in \ref{subsec_photo}). For HD~189733b, a comparably big classical satellite with the maximum possible size of $\approx$640~km is possible. However, the temperature at the orbital distance of this planet is too low to expect the formation of big lava oceans on the surface of the satellite. Maybe a classical satellite orbiting HD~189733b could be a volcanically active body due to tidal interactions with the host planet during its orbital decay. However, this outgassing rates would be difficult to estimate. In any case, as we have shown in sections \ref{subsec_Io} and \ref{subsec_carbon}, a volcanically active body is not expected to outgas carbon.

As for WASP-12b, the planet can host only a much smaller classical satellite in comparison to HD~189733b due to smaller Hill sphere and closer proximity to the star. However, this dynamical estimate does not account for an extreme expansion of the atmosphere of WASP-12b. According to hydrodynamic simulations, this hot Jupiters experiences strong Roche lobe overflow \citep{Bisikalo13,Bisikalo15}, i.e., it fills its whole Roche lobe with material escaping from the upper layers of its atmosphere. Under these conditions, it is difficult to expect the presence of an exomoon: the only possible stabe orbits lie inside an area filled with planetary escaping particles.

\section{Summary and conclusions}
\label{sec_conclusions}

We have considered the influence that rocky evaporating objects can have on optical and UV observations of hot Jupiter systems. We have proposed evaporating lava oceans on the surface of Trojan satellites as a possible plasma source. Outgassed material is ionized and can form a plasma torus which may then leave a signature in the observations, similar to a plasma torus of WASP-12b. We emphasize that the sizes and/or life times of classical moons orbiting hot Jupiters are limited, and in case of WASP-12b the presence of a classical moon can be excluded. The separation between exomoon and hot Jupiter has to be small, depending on the size of the Hill sphere of the planet. Trojan satellites on tadpole orbits near the Lagrangian points $L_4$ and $L_5$, on the other hand, have less severe limitations and present a better option for an outgassing source.

We have shown that Trojan satellites orbiting in 1:1 resonance with the giant exoplanet are stable in both the WASP-12 and HD~189733 systems. We did not find any stable horse-shoe orbits in these systems, however, they are not excluded a priori.

Our conclusions regarding the influence of the photogravitational effects on the stability of small asteroids in systems hosting hot Jupiters were more pessimistic than the results of \citet{Moldovan10}. We have shown that the objects with sizes below several kilometres are destabilized and removed from the system within a short period of time, therefore, there exist a higher lower limit of the size/mass of coorbital satellites under these conditions in comparison to more distant orbits.

We have calculated the stable tadpole orbits at $L_4$ and $L_5$ in both WASP-12 and HD~189733 systems. According to our results, they are distributed over a significant range of possible elongations and inclinations, allowing numerous Trojan satellites to coexist. For two satellites orbiting in $L_4$ and $L_5$, the outgassed material is  believed to form a cloud along the planetary orbit. For a full confidence the simulations similar to the ones performed by \citet{Ben-Jaffel14} and \citet{Mura11} are required, however, assuming a different satellites location and outgassing rates. They will be the focus of a future study. 

There is also another possibility to produce an optically thick cloud around WASP-12b and other hot Jupiters, namely, collision between two or more Trojan asteroids. It has been shown by \citet{Jackson14} that such collisions between medium size bodies in debris discs can lead to the formation of an optically thick cloud which can remain stable for a significant amount of time (up to $\approx$1000 orbital periods of the progenitor, which corresponds to several years for WASP-12b). In this study we did not consider this possibility and did not calculate the collisions probability for the asteroids, but it is clear that some collisions in the remnant of Trojan asteroid swarm still not eliminated by photogravitational effects are possible. This mechanism, however, can not operate in case of two Io-sized Trojan satellites.

For the WASP-12 system, the temperature at the distance of WASP-12b is high enough to melt the surface of the rocky satellites. As a consequence, magma oceans can outgas species like Na, O, O$_2$, SiO, Mg, Ca, Fe, etc., which are rapidly lost to space and ionized by the stellar radiation. This can lead to the formation of a thick constantly replenishing plasma cloud of Mg{\sc ii}, Ca{\sc ii}, and other ions, surrounding the star and absorbing the stellar radiation in corresponding strong spectral lines, which could possibly explain the anomalously low stellar emission observed in the Mg{\sc ii}\,h\&k and Ca{\sc ii}\,H\&K line cores of WASP-12.

If bigger Trojan satellites or an asteroid swarm as an outgassing source are present in the system, WASP-12b is constantly moving in a plasma medium. Since the planet's orbital speed is supersonic, moving through such medium would lead to the formation of a bow shock at a certain distance from the planet, defined by the orbital speed, intrinsic or induced magnetic moment, plasma cloud and stellar wind density, and the planetary wind. The ions of the surmised plasma cloud could not easily penetrate the bow shock and would be piled up in this region, leading to the increased density of the ions at the bow shock and contributing to the early ingress observed in NUV transit observations of WASP-12b.

We have analysed the stability of the Trojan satellites in the HD~189733 system and have shown that, similarly to the case of WASP-12, putative Trojan satellites can orbit on numerous stable orbits, covering the whole stellar disc. Due to lower temperature in this system, the rocky surface of potential Trojan satellites would possess smaller lava oceans compared to WASP-12 Trojans and outgas species like Na, O$_2$, O, and Fe. Absorption features at the wavelengths of specific elements may be searched in UV and optical spectra of the most close-in planets at orbital phases close to that of the $L_4$ and $L_5$ Lagrangian points. However, we emphasize that the mineral atmosphere produced at the equilibrium temperature of HD~189733b can be very sparse and form a very thin plasma cloud, possibly not thick enough to fully absorb the stellar radiation at corresponding spectral lines. For HD~189733b, surface sputtering can be of importance, but this process can not provide outgassing rates comparable to a lava ocean.

Neither for the WASP-12 nor for HD~189733 systems we have found a possibility to produce carbon atoms. We analysed the possibilities to outgas carbon from an Io-type exomoon and have found that this mechanism can not operate for HD~189733 and WASP-12. Also, the Trojan outgassing can not provide sufficient amount of C atoms, leading us to the conclusion that the early ingress observed at the FUV C{\sc ii} doublet for HD~189733b should have an other (e.g., instrumental or planetary) origin.

We surmise that the same mechanism leading to constant absorption of the stellar emission in some lines in NUV can operate also in other exoplanetary systems hosting transiting or grazing extreme hot Jupiters with the equilibrium temperature exceeding 2000~K.

\section*{ACKNOWLEDGMENTS}

This study was carried out with the support by the FWF NFN project S11601-N16 ``Pathways to Habitability: From Disks to Active Stars, Planets and Life" and the related subprojects S11607-N16 and S11608-N16. H.L. thanks the support by FWF project ``Characterizing Stellar and Exoplanetary Environments via Modeling of Lyman-Alpha Transit Observations of Hot Jupiters", P 27256-N27. The authors thank Dr. Yamila Miguel and Dr. Alessandro Mura for their valuable comments on the mineral atmospheres. R.S. thanks the support by the FWF project P23810-N16. The authors thank the anonymous referee for valuable comments.

\bibliography{HJMoons_plasma_torus}

\begin{thebibliography}{}

\bibitem[\protect\citeauthoryear{{Bagenal} \& {Delamere}}{{Bagenal} \&
  {Delamere}}{2011}]{Bagenal11}
{Bagenal} F.,  {Delamere} P.~A.,  2011, Journal of Geophysical Research (Space
  Physics), 116, 5209

\bibitem[\protect\citeauthoryear{{Bakos}, {P{\'a}l}, {Latham}, {Noyes} \&
  {Stefanik}}{{Bakos} et~al.}{2006}]{Bakos06}
{Bakos} G.~{\'A}.,  {P{\'a}l} A.,  {Latham} D.~W.,  {Noyes} R.~W.,
  {Stefanik} R.~P.,  2006, \apjl, 641, L57

\bibitem[\protect\citeauthoryear{{Ballester} \& {Ben-Jaffel}}{{Ballester} \&
  {Ben-Jaffel}}{2015}]{Ballester15}
{Ballester} G.~E.,  {Ben-Jaffel} L.,  2015, \apj, 804, 116

\bibitem[\protect\citeauthoryear{{Barnes} \& {O'Brien}}{{Barnes} \&
  {O'Brien}}{2002}]{Barnes02}
{Barnes} J.~W.,  {O'Brien} D.~P.,  2002, \apj, 575, 1087

\bibitem[\protect\citeauthoryear{{Beaug{\'e}}, {S{\'a}ndor}, {{\'E}rdi} \&
  {S{\"u}li}}{{Beaug{\'e}} et~al.}{2007}]{Beauge07}
{Beaug{\'e}} C.,  {S{\'a}ndor} Z.,  {{\'E}rdi} B.,    {S{\"u}li} {\'A}.,  2007,
  \aap, 463, 359

\bibitem[\protect\citeauthoryear{{Bechter}, {Crepp}, {Ngo}, {Knutson},
  {Batygin}, {Hinkley}, {Muirhead}, {Johnson}, {Howard}, {Montet}, {Matthews}
  \& {Morton}}{{Bechter} et~al.}{2014}]{Bechter14}
{Bechter} E.~B.,  {Crepp} J.~R.,  {Ngo} H.,  {Knutson} H.~A.,  {Batygin} K.,
  {Hinkley} S.,  {Muirhead} P.~S.,  {Johnson} J.~A.,  {Howard} A.~W.,  {Montet}
  B.~T.,  {Matthews} C.~T.,    {Morton} T.~D.,  2014, \apj, 788, 2

\bibitem[\protect\citeauthoryear{{Ben-Jaffel} \& {Ballester}}{{Ben-Jaffel} \&
  {Ballester}}{2014}]{Ben-Jaffel14}
{Ben-Jaffel} L.,  {Ballester} G.~E.,  2014, \apjl, 785, L30

\bibitem[\protect\citeauthoryear{{Bisikalo}, {Kaygorodov}, {Ionov},
  {Shematovich}, {Lammer} \& {Fossati}}{{Bisikalo} et~al.}{2013}]{Bisikalo13}
{Bisikalo} D.,  {Kaygorodov} P.,  {Ionov} D.,  {Shematovich} V.,  {Lammer} H.,
    {Fossati} L.,  2013, \apj, 764, 19

\bibitem[\protect\citeauthoryear{{Bisikalo}, {Kaigorodov} \&
  {Konstantinova}}{{Bisikalo} et~al.}{2015}]{Bisikalo15}
{Bisikalo} D.~V.,  {Kaigorodov} P.~V.,    {Konstantinova} N.~I.,  2015,
  Astronomy Reports, 59, 829

\bibitem[\protect\citeauthoryear{{Boudjada}, {Galopeau}, {Sawas} \&
  {Lammer}}{{Boudjada} et~al.}{2014}]{Boudjada14}
{Boudjada} M.~Y.,  {Galopeau} P.~H.~M.,  {Sawas} S.,    {Lammer} H.,  2014,
  Annales Geophysicae, 32, 1119

\bibitem[\protect\citeauthoryear{{Boyarchuk}, {Bisikalo}, {Kuznetsov} \&
  {Chechetkin}}{{Boyarchuk} et~al.}{2002}]{B2002}
{Boyarchuk} A.~A.,  {Bisikalo} D.~V.,  {Kuznetsov} O.~A.,    {Chechetkin}
  V.~M.,  2002, {Mass transfer in close binary stars}

\bibitem[\protect\citeauthoryear{{Chan}, {Ingemyr}, {Winn}, {Holman},
  {Sanchis-Ojeda}, {Esquerdo} \& {Everett}}{{Chan} et~al.}{2011}]{Chan11}
{Chan} T.,  {Ingemyr} M.,  {Winn} J.~N.,  {Holman} M.~J.,  {Sanchis-Ojeda} R.,
  {Esquerdo} G.,    {Everett} M.,  2011, \aj, 141, 179

\bibitem[\protect\citeauthoryear{{Chrisey}, {Boring}, {Johnson} \&
  {Philipps}}{{Chrisey} et~al.}{1988}]{Chrisey88}
{Chrisey} D.~B.,  {Boring} J.~W.,  {Johnson} R.~E.,    {Philipps} J.~A.,  1988,
  Surface Science, 195, 594

\bibitem[\protect\citeauthoryear{{Crossfield}, {Barman}, {Hansen}, {Tanaka} \&
  {Kodama}}{{Crossfield} et~al.}{2012}]{Crossfield12}
{Crossfield} I.~J.~M.,  {Barman} T.,  {Hansen} B.~M.~S.,  {Tanaka} I.,
  {Kodama} T.,  2012, \apj, 760, 140

\bibitem[\protect\citeauthoryear{{Crouzet}, {McCullough}, {Deming} \&
  {Madhusudhan}}{{Crouzet} et~al.}{2014}]{Crouzet14}
{Crouzet} N.,  {McCullough} P.~R.,  {Deming} D.,    {Madhusudhan} N.,  2014,
  \apj, 795, 166

\bibitem[\protect\citeauthoryear{{Domingos}, {Winter} \& {Yokoyama}}{{Domingos}
  et~al.}{2006}]{Domingos06}
{Domingos} R.~C.,  {Winter} O.~C.,    {Yokoyama} T.,  2006, \mnras, 373, 1227

\bibitem[\protect\citeauthoryear{{Donnison}}{{Donnison}}{2014}]{Donnison14}
{Donnison} J.~R.,  2014, Earth Moon and Planets

\bibitem[\protect\citeauthoryear{{Dvorak}, {Pilat-Lohinger}, {Schwarz} \&
  {Freistetter}}{{Dvorak} et~al.}{2004}]{Dvorak04}
{Dvorak} R.,  {Pilat-Lohinger} E.,  {Schwarz} R.,    {Freistetter} F.,  2004,
  \aap, 426, L37

\bibitem[\protect\citeauthoryear{{Erkaev}, {Lammer}, {Odert}, {Kulikov},
  {Kislyakova}, {Khodachenko}, {G{\"u}del}, {Hanslmeier} \& {Biernat}}{{Erkaev}
  et~al.}{2013}]{Erkaev13}
{Erkaev} N.~V.,  {Lammer} H.,  {Odert} P.,  {Kulikov} Y.~N.,  {Kislyakova}
  K.~G.,  {Khodachenko} M.~L.,  {G{\"u}del} M.,  {Hanslmeier} A.,    {Biernat}
  H.,  2013, Astrobiology, 13, 1011

\bibitem[\protect\citeauthoryear{{Fegley} \& {Zolotov}}{{Fegley} \&
  {Zolotov}}{2000}]{Fegley00}
{Fegley} B.,  {Zolotov} M.~Y.,  2000, \icarus, 148, 193

\bibitem[\protect\citeauthoryear{{Figueira}, {Oshagh}, {Adibekyan} \&
  {Santos}}{{Figueira} et~al.}{2014}]{Figueira2014}
{Figueira} P.,  {Oshagh} M.,  {Adibekyan} V.~Z.,    {Santos} N.~C.,  2014,
  \aap, 572, A51

\bibitem[\protect\citeauthoryear{{Fossati}, {Ayres}, {Haswell}, {Bohlender},
  {Kochukhov} \& {Fl{\"o}er}}{{Fossati} et~al.}{2013}]{Fossati13}
{Fossati} L.,  {Ayres} T.~R.,  {Haswell} C.~A.,  {Bohlender} D.,  {Kochukhov}
  O.,    {Fl{\"o}er} L.,  2013, \apjl, 766, L20

\bibitem[\protect\citeauthoryear{{Fossati}, {Ayres}, {Haswell}, {Jenkins},
  {Bisikalo}, {Bohlender}, {Fl{\"o}er} \& {Kochukhov}}{{Fossati}
  et~al.}{2014}]{Fossati14}
{Fossati} L.,  {Ayres} T.~R.,  {Haswell} C.~A.,  {Jenkins} J.~S.,  {Bisikalo}
  D.,  {Bohlender} D.,  {Fl{\"o}er} L.,    {Kochukhov} O.,  2014, \apss, 354,
  21

\bibitem[\protect\citeauthoryear{{Fossati}, {Haswell}, {Froning}, {Hebb},
  {Holmes}, {Kolb}, {Helling}, {Carter}, {Wheatley}, {Collier Cameron},
  {Loeillet}, {Pollacco}, {Street}, {Stempels}, {Simpson}, {Udry}, {Joshi} \&
  {et al.}}{{Fossati} et~al.}{2010}]{Fossati10}
{Fossati} L.,  {Haswell} C.~A.,  {Froning} C.~S.,  {Hebb} L.,  {Holmes} S.,
  {Kolb} U.,  {Helling} C.,  {Carter} A.,  {Wheatley} P.,  {Collier Cameron}
  A.,  {Loeillet} B.,  {Pollacco} D.,  {Street} R.,  {Stempels} H.~C.,
  {Simpson} E.,  {Udry} S.,  {Joshi} Y.~C.,    {et al.} 2010, \apjl, 714, L222

\bibitem[\protect\citeauthoryear{{Fossati}, {Ingrassia} \& {Lanza}}{{Fossati}
  et~al.}{2015}]{Fossati15}
{Fossati} L.,  {Ingrassia} S.,    {Lanza} A.~F.,  2015, \apjl, 812, L35

\bibitem[\protect\citeauthoryear{{Froeschl{\'e}}, {Gonczi} \&
  {Lega}}{{Froeschl{\'e}} et~al.}{1997}]{Froeschle97}
{Froeschl{\'e}} C.,  {Gonczi} R.,    {Lega} E.,  1997, \planss, 45, 881

\bibitem[\protect\citeauthoryear{{Galiazzo} \& {Schwarz}}{{Galiazzo} \&
  {Schwarz}}{2014}]{Galiazzo14}
{Galiazzo} M.~A.,  {Schwarz} R.,  2014, \mnras, 445, 3999

\bibitem[\protect\citeauthoryear{{Guenther}, {Cabrera}, {Erikson}, {Fridlund},
  {Lammer}, {Mura}, {Rauer}, {Schneider}, {Tulej}, {von Paris} \&
  {Wurz}}{{Guenther} et~al.}{2011}]{Guenther11}
{Guenther} E.~W.,  {Cabrera} J.,  {Erikson} A.,  {Fridlund} M.,  {Lammer} H.,
  {Mura} A.,  {Rauer} H.,  {Schneider} J.,  {Tulej} M.,  {von Paris} P.,
  {Wurz} P.,  2011, \aap, 525, A24

\bibitem[\protect\citeauthoryear{{Hartman}}{{Hartman}}{2010}]{Hartman2010}
{Hartman} J.~D.,  2010, \apjl, 717, L138

\bibitem[\protect\citeauthoryear{{Haswell}, {Fossati}, {Ayres}, {France},
  {Froning}, {Holmes}, {Kolb}, {Busuttil}, {Street}, {Hebb}, {Collier Cameron},
  {Enoch}, {Burwitz}, {Rodriguez}, {West}, {Pollacco}, {Wheatley} \&
  {Carter}}{{Haswell} et~al.}{2012}]{Haswell12}
{Haswell} C.~A.,  {Fossati} L.,  {Ayres} T.,  {France} K.,  {Froning} C.~S.,
  {Holmes} S.,  {Kolb} U.~C.,  {Busuttil} R.,  {Street} R.~A.,  {Hebb} L.,
  {Collier Cameron} A.,  {Enoch} B.,  {Burwitz} V.,  {Rodriguez} J.,  {West}
  R.~G.,  {Pollacco} D.,  {Wheatley} P.~J.,    {Carter} A.,  2012, \apj, 760,
  79

\bibitem[\protect\citeauthoryear{{Heller}}{{Heller}}{2014}]{Heller14b}
{Heller} R.,  2014, \apj, 787, 14

\bibitem[\protect\citeauthoryear{{Ito}, {Ikoma}, {Kawahara}, {Nagahara},
  {Kawashima} \& {Nakamoto}}{{Ito} et~al.}{2015}]{Ito15}
{Ito} Y.,  {Ikoma} M.,  {Kawahara} H.,  {Nagahara} H.,  {Kawashima} Y.,
  {Nakamoto} T.,  2015, \apj, 801, 144

\bibitem[\protect\citeauthoryear{{Jackson}, {Wyatt}, {Bonsor} \&
  {Veras}}{{Jackson} et~al.}{2014}]{Jackson14}
{Jackson} A.~P.,  {Wyatt} M.~C.,  {Bonsor} A.,    {Veras} D.,  2014, \mnras,
  440, 3757

\bibitem[\protect\citeauthoryear{{Jewitt} \& {Haghighipour}}{{Jewitt} \&
  {Haghighipour}}{2007}]{Jewitt07}
{Jewitt} D.,  {Haghighipour} N.,  2007, \araa, 45, 261

\bibitem[\protect\citeauthoryear{{Jewitt}, {Trujillo} \& {Luu}}{{Jewitt}
  et~al.}{2000}]{Jewitt00}
{Jewitt} D.~C.,  {Trujillo} C.~A.,    {Luu} J.~X.,  2000, \aj, 120, 1140

\bibitem[\protect\citeauthoryear{{Kley} \& {Nelson}}{{Kley} \&
  {Nelson}}{2008}]{Kley08}
{Kley} W.,  {Nelson} R.~P.,  2008, \aap, 486, 617

\bibitem[\protect\citeauthoryear{{Kozai}}{{Kozai}}{1962}]{Kozai62}
{Kozai} Y.,  1962, \aj, 67, 591

\bibitem[\protect\citeauthoryear{{Lammer}, {Schiefer}, {Juvan}, {Odert},
  {Erkaev}, {Weber}, {Kislyakova}, {G{\"u}del}, {Kirchengast} \&
  {Hanslmeier}}{{Lammer} et~al.}{2014}]{Lammer14}
{Lammer} H.,  {Schiefer} S.-C.,  {Juvan} I.,  {Odert} P.,  {Erkaev} N.~V.,
  {Weber} C.,  {Kislyakova} K.~G.,  {G{\"u}del} M.,  {Kirchengast} G.,
  {Hanslmeier} A.,  2014, Origins of Life and Evolution of the Biosphere, 44,
  239

\bibitem[\protect\citeauthoryear{{Lanza}}{{Lanza}}{2014}]{Lanza2014}
{Lanza} A.~F.,  2014, \aap, 572, L6

\bibitem[\protect\citeauthoryear{{Lidov}}{{Lidov}}{1962}]{Lidov62}
{Lidov} M.~L.,  1962, \planss, 9, 719

\bibitem[\protect\citeauthoryear{{Lin}, {Bodenheimer} \& {Richardson}}{{Lin}
  et~al.}{1996}]{Lin96}
{Lin} D.~N.~C.,  {Bodenheimer} P.,    {Richardson} D.~C.,  1996, \nat, 380, 606

\bibitem[\protect\citeauthoryear{{Llama}, {Wood}, {Jardine}, {Vidotto},
  {Helling}, {Fossati} \& {Haswell}}{{Llama} et~al.}{2011}]{Llama2011}
{Llama} J.,  {Wood} K.,  {Jardine} M.,  {Vidotto} A.~A.,  {Helling} C.,
  {Fossati} L.,    {Haswell} C.~A.,  2011, \mnras, 416, L41

\bibitem[\protect\citeauthoryear{{Maciejewski}, {Dimitrov}, {Seeliger},
  {Raetz}, {Bukowiecki}, {Kitze}, {Errmann}, {Nowak}, {Niedzielski}, {Popov},
  {Marka}, {Go{\'z}dziewski}, {Neuh{\"a}user}, {Ohlert} \& {et
  al.}}{{Maciejewski} et~al.}{2013}]{Maciejewski13}
{Maciejewski} G.,  {Dimitrov} D.,  {Seeliger} M.,  {Raetz} S.,  {Bukowiecki}
  {\L}.,  {Kitze} M.,  {Errmann} R.,  {Nowak} G.,  {Niedzielski} A.,  {Popov}
  V.,  {Marka} C.,  {Go{\'z}dziewski} K.,  {Neuh{\"a}user} R.,  {Ohlert} J.,
  {et al.} 2013, \aap, 551, A108

\bibitem[\protect\citeauthoryear{{Markellos}, {Roy}, {Velgakis} \&
  {Kanavos}}{{Markellos} et~al.}{2000}]{Markellos00}
{Markellos} V.~V.,  {Roy} A.~E.,  {Velgakis} M.~J.,    {Kanavos} S.~S.,  2000,
  \apss, 271, 293

\bibitem[\protect\citeauthoryear{{McEwen}, {Keszthelyi}, {Spencer}, {Schubert},
  {Matson}, {Lopes-Gautier}, {Klaasen}, {Johnson}, {Head}, {Geissler},
  {Fagents}, {Davies}, {Carr}, {Breneman} \& {Belton}}{{McEwen}
  et~al.}{1998}]{McEwen98}
{McEwen} A.~S.,  {Keszthelyi} L.,  {Spencer} J.~R.,  {Schubert} G.,  {Matson}
  D.~L.,  {Lopes-Gautier} R.,  {Klaasen} K.~P.,  {Johnson} T.~V.,  {Head}
  J.~W.,  {Geissler} P.,  {Fagents} S.,  {Davies} A.~G.,  {Carr} M.~H.,
  {Breneman} H.~H.,    {Belton} M.~J.~S.,  1998, Science, 281, 87

\bibitem[\protect\citeauthoryear{{Miguel}, {Kaltenegger}, {Fegley} \&
  {Schaefer}}{{Miguel} et~al.}{2011}]{Miguel11}
{Miguel} Y.,  {Kaltenegger} L.,  {Fegley} B.,    {Schaefer} L.,  2011, \apjl,
  742, L19

\bibitem[\protect\citeauthoryear{{Moldovan}, {Matthews}, {Gladman}, {Bottke} \&
  {Vokrouhlick{\'y}}}{{Moldovan} et~al.}{2010}]{Moldovan10}
{Moldovan} R.,  {Matthews} J.~M.,  {Gladman} B.,  {Bottke} W.~F.,
  {Vokrouhlick{\'y}} D.,  2010, \apj, 716, 315

\bibitem[\protect\citeauthoryear{{Morbidelli}, {Levison}, {Tsiganis} \&
  {Gomes}}{{Morbidelli} et~al.}{2005}]{Morbidelli05}
{Morbidelli} A.,  {Levison} H.~F.,  {Tsiganis} K.,    {Gomes} R.,  2005, \nat,
  435, 462

\bibitem[\protect\citeauthoryear{{Moses}, {Zolotov} \& {Fegley}}{{Moses}
  et~al.}{2002}]{Moses02}
{Moses} J.~I.,  {Zolotov} M.~Y.,    {Fegley} B.,  2002, \icarus, 156, 76

\bibitem[\protect\citeauthoryear{{Mura}, {Wurz}, {Schneider}, {Lammer},
  {Grie{\ss}meier}, {Khodachenko}, {Weingrill}, {Guenther}, {Cabrera},
  {Erikson}, {Fridlund}, {Milillo}, {Rauer} \& {von Paris}}{{Mura}
  et~al.}{2011}]{Mura11}
{Mura} A.,  {Wurz} P.,  {Schneider} J.,  {Lammer} H.,  {Grie{\ss}meier} J.-M.,
  {Khodachenko} M.~L.,  {Weingrill} J.,  {Guenther} E.,  {Cabrera} J.,
  {Erikson} A.,  {Fridlund} M.,  {Milillo} A.,  {Rauer} H.,    {von Paris} P.,
  2011, \icarus, 211, 1

\bibitem[\protect\citeauthoryear{{Namouni}}{{Namouni}}{2010}]{Namouni10}
{Namouni} F.,  2010, \apjl, 719, L145

\bibitem[\protect\citeauthoryear{{Naoz}, {Farr}, {Lithwick}, {Rasio} \&
  {Teyssandier}}{{Naoz} et~al.}{2011}]{Naoz11}
{Naoz} S.,  {Farr} W.~M.,  {Lithwick} Y.,  {Rasio} F.~A.,    {Teyssandier} J.,
  2011, \nat, 473, 187

\bibitem[\protect\citeauthoryear{{Nesvorn{\'y}} \&
  {Vokrouhlick{\'y}}}{{Nesvorn{\'y}} \& {Vokrouhlick{\'y}}}{2009}]{Nesvorny09}
{Nesvorn{\'y}} D.,  {Vokrouhlick{\'y}} D.,  2009, \aj, 137, 5003

\bibitem[\protect\citeauthoryear{{Peale}, {Cassen} \& {Reynolds}}{{Peale}
  et~al.}{1979}]{Peale79}
{Peale} S.~J.,  {Cassen} P.,    {Reynolds} R.~T.,  1979, Science, 203, 892

\bibitem[\protect\citeauthoryear{{Placek}, {Knuth}, {Angerhausen} \&
  {Jenkins}}{{Placek} et~al.}{2015}]{Placek15}
{Placek} B.,  {Knuth} K.~H.,  {Angerhausen} D.,    {Jenkins} J.~M.,  2015,
  \apj, 814, 147

\bibitem[\protect\citeauthoryear{{Podlewska-Gaca} \&
  {Szuszkiewicz}}{{Podlewska-Gaca} \& {Szuszkiewicz}}{2014}]{Podlewska-Gaca14}
{Podlewska-Gaca} E.,  {Szuszkiewicz} E.,  2014, \mnras, 438, 2538

\bibitem[\protect\citeauthoryear{{Poppenhaeger}, {Schmitt} \&
  {Wolk}}{{Poppenhaeger} et~al.}{2013}]{Poppenhaeger13}
{Poppenhaeger} K.,  {Schmitt} J.~H.~M.~M.,    {Wolk} S.~J.,  2013, \apj, 773,
  62

\bibitem[\protect\citeauthoryear{{Postberg}, {Kempf}, {Srama}, {Green},
  {Hillier}, {McBride} \& {Gr{\"u}n}}{{Postberg} et~al.}{2006}]{Postberg06}
{Postberg} F.,  {Kempf} S.,  {Srama} R.,  {Green} S.~F.,  {Hillier} J.~K.,
  {McBride} N.,    {Gr{\"u}n} E.,  2006, \icarus, 183, 122

\bibitem[\protect\citeauthoryear{{Rasio} \& {Ford}}{{Rasio} \&
  {Ford}}{1996}]{Rasio96}
{Rasio} F.~A.,  {Ford} E.~B.,  1996, Science, 274, 954

\bibitem[\protect\citeauthoryear{{Schaefer} \& {Fegley}}{{Schaefer} \&
  {Fegley}}{2005}]{Schaefer05}
{Schaefer} L.,  {Fegley} B.,  2005, \icarus, 173, 454

\bibitem[\protect\citeauthoryear{{Schaefer} \& {Fegley}}{{Schaefer} \&
  {Fegley}}{2009}]{Schaefer09}
{Schaefer} L.,  {Fegley} B.,  2009, \apjl, 703, L113

\bibitem[\protect\citeauthoryear{{Schuerman}}{{Schuerman}}{1980}]{Schuerman80}
{Schuerman} D.~W.,  1980, \apj, 238, 337

\bibitem[\protect\citeauthoryear{{Schwarz} \& {Dvorak}}{{Schwarz} \&
  {Dvorak}}{2012}]{Schwarz12}
{Schwarz} R.,  {Dvorak} R.,  2012, Celestial Mechanics and Dynamical Astronomy,
  113, 23

\bibitem[\protect\citeauthoryear{{Shaikhislamov}, {Khodachenko}, {Sasunov},
  {Lammer}, {Kislyakova} \& {Erkaev}}{{Shaikhislamov}
  et~al.}{2014}]{Shaikhislamov14}
{Shaikhislamov} I.~F.,  {Khodachenko} M.~L.,  {Sasunov} Y.~L.,  {Lammer} H.,
  {Kislyakova} K.~G.,    {Erkaev} N.~V.,  2014, \apj, 795, 132

\bibitem[\protect\citeauthoryear{{Simon}, {Szab{\'o}}, {Kiss}, {Fortier} \&
  {Benz}}{{Simon} et~al.}{2015}]{Simon15}
{Simon} A.~E.,  {Szab{\'o}} G.~M.,  {Kiss} L.~L.,  {Fortier} A.,    {Benz} W.,
  2015, \pasp, 127, 1084

\bibitem[\protect\citeauthoryear{{Sing}, {Lecavelier des Etangs}, {Fortney},
  {Burrows}, {Pont}, {Wakeford}, {Ballester}, {Nikolov}, {Henry}, {Aigrain},
  {Deming} \& {et al.}}{{Sing} et~al.}{2013}]{Sing13}
{Sing} D.~K.,  {Lecavelier des Etangs} A.,  {Fortney} J.~J.,  {Burrows} A.~S.,
  {Pont} F.,  {Wakeford} H.~R.,  {Ballester} G.~E.,  {Nikolov} N.,  {Henry}
  G.~W.,  {Aigrain} S.,  {Deming} D.,    {et al.} 2013, \mnras, 436, 2956

\bibitem[\protect\citeauthoryear{{Spalding}, {Batygin} \& {Adams}}{{Spalding}
  et~al.}{2016}]{Spalding16}
{Spalding} C.,  {Batygin} K.,    {Adams} F.~C.,  2016, \apj, 817, 18

\bibitem[\protect\citeauthoryear{{Thomas}, {Bagenal}, {Hill} \&
  {Wilson}}{{Thomas} et~al.}{2004}]{Thomas04}
{Thomas} N.,  {Bagenal} F.,  {Hill} T.~W.,    {Wilson} J.~K.,  2004, {The Io
  neutral clouds and plasma torus}.
pp 561--591

\bibitem[\protect\citeauthoryear{{Veras}, {Eggl} \& {G{\"a}nsicke}}{{Veras}
  et~al.}{2015}]{Veras15}
{Veras} D.,  {Eggl} S.,    {G{\"a}nsicke} B.~T.,  2015, \mnras, 451, 2814

\bibitem[\protect\citeauthoryear{{Vidotto}, {Jardine} \& {Helling}}{{Vidotto}
  et~al.}{2010}]{Vidotto10}
{Vidotto} A.~A.,  {Jardine} M.,    {Helling} C.,  2010, \apjl, 722, L168

\bibitem[\protect\citeauthoryear{{Weidner} \& {Horne}}{{Weidner} \&
  {Horne}}{2010}]{Weidner10}
{Weidner} C.,  {Horne} K.,  2010, \aap, 521, A76

\bibitem[\protect\citeauthoryear{{Wong} \& {Smyth}}{{Wong} \&
  {Smyth}}{2000}]{Wong00}
{Wong} M.~C.,  {Smyth} W.~H.,  2000, \icarus, 146, 60

\end{thebibliography}
\bibliographystyle{mn2e}

\end{document}